\documentclass[%
prd,
nofootinbib,
superscriptaddress
]{revtex4}
\usepackage{amsmath,amssymb,bm,float,mathrsfs}
\usepackage[dvipdfmx]{graphicx}
\usepackage{dcolumn}
\usepackage{ulem}
\usepackage{color}
\usepackage{hyperref}
\usepackage{comment}
\usepackage{framed}
\usepackage{url}
\hypersetup{
    colorlinks=true,
    citecolor=cyan,
}

  \newcommand{\beq}{\begin{equation}}
  \newcommand{\eeq}{\end{equation}}
  \newcommand{\al}[1]{\begin{align} #1 \end{align}}
  \newcommand{\bi}{\begin{itemize}}
  \newcommand{\ei}{\end{itemize}}
  
  \def\dd{\mathrm{d}}

  \def\pd{\partial}
  \newcommand{\ave}[1]{\left\langle #1 \right\rangle}

\begin{document}

\title{Second-order peculiar velocity field as a novel probe of scalar-tensor theories}


\author{Daisuke Yamauchi}
\email[Email: ]{yamauchi"at"jindai.jp}
\affiliation{
Faculty of Engineering, Kanagawa University, Kanagawa, 221-8686, Japan
}

\author{Naonori S. Sugiyama}
\email[Email: ]{nao.s.sugiyama"at"gmail.com}
\affiliation{
National Astronomical Observatory of Japan, Mitaka, Tokyo 181-8588, Japan
}

\begin{abstract}
We investigate the galaxy bispectrum induced by the nonlinear gravitational evolution as a possible probe to constrain
degenerate higher-order scalar tensor (DHOST) theories. We find that the signal obtained from the leading kernel of second-order density fluctuations is partially hidden by the uncertainty in the nonlinear galaxy bias, and that the kernel of second-order velocity fields instead provides unbiased information on the modification of gravity theory. Based on this fact, we propose new phenomenological time-dependent functions, written as a combination of the coefficients of the second-order kernels, which is expected to trace the higher-order growth history. We then present approximate expressions for these variables in terms of parameters that characterize the DHOST theories. We also show that the resultant formulae provides new constraints on the parameter space of the DHOST theories.

\end{abstract}


\maketitle

\section{Introduction}
\label{sec:Introduction}

The accelerating cosmic expansion could arise due to a modification of general relativity on cosmological scales. Various theoretical scenarios have been proposed in the literature and should be carefully compared with observational data. Among the cosmological observational data, measuring the growth history of density fluctuations is a powerful tool to test the nature of dark energy and the modification of the theory of gravity responsible for the present cosmic acceleration. In order to efficiently compare the observational data with theoretical predictions, it is convenient to consider phenomenological parameters. A minimal approach to test the theory of gravity from measurements of the growth rate of large-scale structure (LSS) is to introduce the gravitational growth index $\gamma$. This parameter is defined by the logarithmic derivative of the growth rate $f$ with respect to the fractional parameter of the non-relativistic matter density $\Omega_{\rm m}$ as~\cite{Linder:2007hg} 
\al{
	&\gamma :=\frac{\dd\ln f}{\dd\ln\Omega_{\rm m}}
	\,.\label{eq:growth index def}
}
In the standard cosmological model responsible for the present cosmic acceleration, i.e., $\Lambda$ cold dark matter ($\Lambda$CDM) model with general relativity, one shows $\gamma$ to be nearly constant at $\gamma\approx 0.545$. Although constraints on the growth index have been reported in the literature~\cite{Grieb:2016uuo,Sanchez:2016sas,Gil-Marin:2018cgo,Zhao:2018gvb},
there is no evidence that they deviate from the value predicted by the standard $\Lambda$CDM model.
Nevertheless, further exploration of the cosmological model landscape requires introducing additional parameters to capture modifications to the theory of gravity.

A possible candidate is a second-order index developed in~\cite{Yamauchi:2017ibz} (see also \cite{Namikawa:2018erh}).
Modifications to the theory of gravity typically alter the clustering properties of LSS. Thus, the quasi-nonlinear growth of LSS can provide new insights into the theory of gravity that would not be imprinted in the growth index of linear perturbation theory. By observing the higher-order correlation function of LSS such as the galaxy bispectrum~\cite{Gil-Marin:2016wya,Slepian:2016kfz,Pearson:2017wtw,Sugiyama:2018yzo,Sugiyama:2020uil}, we can explore the quasi-nonlinear growth of LSS described by the nonlinear kernel of density fluctuations. We define the second-order index as the logarithmic derivative of the time-evolving coefficient $\zeta =\kappa,\lambda$ in the second-order kernel with respect to $\Omega_{\rm m}$ as
\al{
	&\xi_\zeta :=\frac{\dd\ln\zeta}{\dd\ln\Omega_{\rm m}}
	\,.\label{eq:second-order index def}
}
We expect that second-order indexes can deliver new information on the modification of gravity theories and break the degeneracy between cosmological parameters. Therefore, the purpose of this paper is to revisit the second-order indexes and apply them to a broad class of scalar-tensor gravity theories called degenerate higher-order scalar-tensor theories (DHOST)~\cite{Langlois:2015cwa,Crisostomi:2016czh,Achour:2016rkg,BenAchour:2016fzp} (for a review, see \cite{Langlois:2018dxi,Kobayashi:2019hrl}). However, as will be shown in the next section, 
the kernel of the second-order density fluctuation always appears together with the nonlinear galaxy bias functions, making it difficult to directly determine one of the second-order indexes by measuring the galaxy bispectrum. In other words, the features of the second-order indexes are partially hidden by the uncertainty of the nonlinear galaxy bias. To avoid this problem, in this paper,
we propose new phenomenological time-dependent functions $E_{\rm f}=\Omega_{\rm m}^{\xi_{\rm f}}$, $E_{\rm s}=\Omega_{\rm m}^{\xi_{\rm s}}$ and $E_{\rm t}=\Omega_{\rm m}^{\xi_{\rm t}}$, focusing on the contributions from the second-order peculiar velocity field. We expect that $E_{\rm s}$ and $E_{\rm t}$ can be used to constrain modifications to the theory of gravity without any observational uncertainties.
In order to compare observational data with theoretical predictions, we develop a formalism describing the evolution equations for the second-order perturbations and derive their explicit expressions as functions of the effective-field-theory (EFT) parameters describing the DHOST theories. 

This paper is organized as follows.
In Sec.~\ref{sec:Galaxy bisepctrum in redshift space}, we first give
the basic equations for the galaxy bispectrum in redshift space
and discuss the observational difficulties in the presence of the galaxy nonlinear bias.
In Sec.~\ref{sec:Nonlinear gravitational growth in degenerate higher-order scalar-tensor theories},
following \cite{Hirano:2020dom} we show the effective Lagrangian describing the DHOST theories
and derive the evolution equations for the first- and second-order density fluctuations.
We then evaluate the growth index and the second-order index and give
their approximate expressions in Sec.~\ref{sec:Approximate expression}.
We then apply the resultant formulae to the shift-symmetric DHOST cosmology 
as an application. 
Finally, Sec.~\ref{sec:Summary} is devoted to a summary and conclusion.

\section{Galaxy bispectrum in redshift space}
\label{sec:Galaxy bisepctrum in redshift space}

In order to derive the galaxy density fluctuation in redshift space as an observable of galaxy redshift surveys,
we first need to describe the matter density field $\delta (t,{\bm x})$ and the peculiar velocity field ${\bm v}(t,{\bm x})$. For the pressureless nonrelativistic matter, the evolution equation for linear density fluctuations does not depend on the wave number, even when gravity theory is described by a broad class of modified gravity theories, in particular the DHOST theories, 
as will be discussed later.
Hence, the time-dependence of the linear density field can be expressed independently of the wave number,
namely $\delta(t,{\bm k})=D_+(t)\delta_{\rm L}({\bm k})$, where $D_+(t)$ and $\delta_{\rm L}({\bm k})$ denote the linear growth and the initial density fluctuation. On the other hand, the linear velocity divergence field, $\theta(t,{\bm x}):=-\pd_i v^i(t,{\bm x})/aH$ is written in terms of the logarithmic time derivative of the linear matter density fluctuation through the continuity equation, $\theta(t,{\bm k})=f(t)D_+(t)\delta_{\rm L}({\bm k})$ with $f(t):=\dd\ln D_+/\dd\ln a$. The Fourier transform of the density field and the velocity divergence field are formally expanded in terms of the initial density field $\delta_{\rm L}({\bm k})$ as
\al{
	&\delta (t,{\bm k})
		=D_+(t)\delta_{\rm L}({\bm k})+D_+^2(t)\int\frac{\dd^3{\bm p}_1\dd{\bm p}_2}{(2\pi)^3}
			\delta_{\rm D}^3({\bm k}-{\bm p}_1-{\bm p}_2)F_2({\bm p}_1,{\bm p}_2;t)
			\delta_{\rm L}({\bm p}_1)\delta_{\rm L}({\bm p}_2)+\cdots
	\,,\label{eq:delta expansion}\\
	&\theta (t,{\bm k})
		=f(t)\biggl[
			D_+(t)\delta_{\rm L}({\bm k})+D_+^2(t)\int\frac{\dd^3{\bm p}_1\dd{\bm p}_2}{(2\pi)^3}
				\delta_{\rm D}^3({\bm k}-{\bm p}_1-{\bm p}_2)G_2({\bm p}_1,{\bm p}_2;t)
				\delta_{\rm L}({\bm p}_1)\delta_{\rm L}({\bm p}_2)
				+\cdots
			\biggr]
	\,,
}
where $a$ is the cosmic scale factor, $H=\dot a/a$ with dot being the derivative with respect to the cosmic time.

Since the density field is indirectly related to the observables of large-scale structure, the relation between them is needed. We assume that the galaxy density fluctuation $\delta_{\rm g}(t,{\bm x})$ up to the second-order can be written as the combination of the linear bias $b_1(t)$, the second-order bias $b_2(t)$, and tidal bias $b_{s^2}(t)$ (see e.g. \cite{Desjacques:2016bnm}):
\al{
	\delta_{\rm g}(t,{\bm x})=b_1(t)\delta(t,{\bm x})+\frac{1}{2}b_2(t)\delta^2(t,{\bm x})
					+b_{s^2}(t)\biggl\{
						\biggl[\frac{\pd_i\pd_j}{\pd^2}\delta (t,{\bm x})\biggr]^2-\frac{1}{3}\delta^2 (t,{\bm x})
					\biggr\}+\cdots
	\,.
}
In redshift space, the radial position of galaxies is given by the observed radial component of its relative velocity to an observer. The peculiar velocity field of the underlying matter density distorts the observed position of the galaxy
along the line-of-sight.
The mapping of a galaxy from its position ${\bm x}$ in real space to its position ${\bm s}$ in redshift space along the line-of-sight direction $\widehat{\bm n}$ is expressed as
\al{
	{\bm s}(t,{\bm x})={\bm x}+\frac{{\bm v}(t,{\bm x})\cdot\widehat{\bm n}}{aH}\widehat{\bm n}
	\,.
}
We then obtain the Fourier component of the galaxy density contrast in redshift space $\delta_{{\rm g},s}(t,{\bm k})$ as
\al{
	\delta_{{\rm g},s}(t,{\bm k})
		=Z_1({\bm k};t)D_+(t)\delta_{\rm L}({\bm k})
			+D_+^2(t)\int\frac{\dd^3{\bm p}_1\dd^3{\bm p}_2}{(2\pi)^3}
				\delta_{\rm D}^3({\bm k}-{\bm p}_1-{\bm p}_2)
				Z_2({\bm p}_1,{\bm p}_2;t)\delta_{\rm L}({\bm p}_1)\delta_{\rm L}({\bm p}_2)
			+\cdots
	\,.
}
Here, the linear- and second-order perturbative kernels are defined as \cite{Scoccimarro:1999ed}
\al{
	&Z_1({\bm p})=b_1+f(\widehat{\bm p}\cdot\widehat{\bm n})^2
	\,,\\
	&Z_2({\bm p}_1,{\bm p}_2)
		=b_1F_2({\bm p}_1,{\bm p}_2)+f(\widehat{\bm k}\cdot\widehat{\bm n})^2 G_2({\bm p}_1,{\bm p}_2)
			+\frac{f({\bm k}\cdot\widehat{\bm n})}{2}
				\Bigl[
					\left(\widehat{\bm p}_1\cdot\widehat{\bm n}\right)Z_1({\bm p}_2)
					+\left(\widehat{\bm p}_2\cdot\widehat{\bm n}\right)Z_1({\bm p}_1)
				\Bigr]
	\notag\\
	&\quad\quad\quad\quad\quad\quad
			+\frac{1}{2}b_2
			+b_{s^2}{\cal T}({\bm p}_1,{\bm p}_2)
	\,,
}
where $\widehat{\bm p}\equiv{\bm p}/p$, 
${\bm k}={\bm p}_1+{\bm p}_2$ and ${\cal T}({\bm p}_1,{\bm p}_2)$ represents the scale-dependent function 
corresponding to the tidal force:
\al{
	{\cal T}({\bm p}_1,{\bm p}_2)=\left(\widehat{\bm p}_1\cdot\widehat{\bm p}_2\right)^2-\frac{1}{3}
	\,.\label{eq:tidal}
}
Assuming that the initial density field $\delta_{\rm L}({\bm k})$ obeys the Gaussian statistics with the power spectrum $P_{\rm L}(k)$ defined through $\ave{\delta_{\rm L}({\bm k})\delta_{\rm L}({\bm k}^\prime)}=(2\pi)^3\delta_{\rm D}^3({\bm k}+{\bm k}^\prime)P_{\rm L}(k)$,
the power spectrum and bispectrum of galaxy fluctuations in redshift space at the leading order of perturbation are given by
\al{
	&P({\bm k};t)
		=Z_1^2({\bm k};t)D_+^2(t)P_{\rm L}(k)
	\,,\label{eq:galaxy power spectrum}\\
	&B({\bm k}_1,{\bm k}_2,{\bm k}_3;t)
		=2Z_1({\bm k}_1;t)Z_1({\bm k}_2;t)Z_2({\bm k}_1,{\bm k}_2;t)D_+^4(t)P_{\rm L}(k_1)P_{\rm L}(k_2)
			+(2\ \text{perms})
	\,,\label{eq:galaxy bispectrum}
}
where ${\bm k}_1+{\bm k}_2+{\bm k}_3={\bm 0}$.

The modification of gravity theory alters the clustering properties of nonlinear structures and peculiar velocity fields. In particular, the time-dependence of the second-order perturbative kernels $F_2$ and $G_2$ yields a powerful probe of modified gravity theories. In this paper, we only focus on the DHOST theories, while there is a wide variety of gravity theories that yield different signatures to the nonlinear kernels. In the case of the type-I DHOST theories, the second-order kernels can be written in the form \cite{Takushima:2013foa,Takushima:2015iha,Crisostomi:2019vhj,Lewandowski:2019txi,Hirano:2020dom}:
\al{
	&F_2({\bm p}_1,{\bm p}_2;t)=\kappa (t)\alpha_{\rm s}({\bm p}_1,{\bm p}_2)-\frac{2}{7}\lambda (t)\gamma({\bm p}_1,\bm p_2)
	\,,\label{eq:F_2}\\
	&G_2({\bm p}_1,{\bm p}_2;t)=\kappa_\theta (t)\alpha_{\rm s}({\bm p}_1,{\bm p}_2)
						-\frac{4}{7}\lambda_\theta (t)\gamma ({\bm p}_1,{\bm p}_2)
	\,,
}
where $\alpha_{\rm s}({\bm p}_1,{\bm p}_2)$ and $\gamma ({\bm p}_1,{\bm p}_2)$ represent
the kernel characterizing the second-order mode coupling
\al{
	&\alpha_{\rm s}({\bm p}_1,{\bm p}_2)
		=1+\frac{1}{2}(\widehat{\bm p}_1\cdot\widehat{\bm p}_2)
			\left(\frac{p_1}{p_2}+\frac{p_2}{p_1}\right)
	\,,\label{eq:mode coupling alpha_s def}\\
	&\gamma ({\bm p}_1,{\bm p}_2)
		=1-(\widehat{\bm p}_1\cdot\widehat{\bm p}_2)^2
	\,.\label{eq:mode coupling gamma def}
}
Here, when assuming that the matter sector is minimally coupled with gravity, 
the continuity and Euler equations for the matter give the relation between
the coefficients of $F_2$ and $G_2$ kernels as
\al{
	\kappa_\theta &= 2\kappa - 1 + \frac{\dot\kappa}{fH}
	\,,\label{eq:kappa kappa_theta}\\
	\lambda_\theta &= \lambda +\frac{\dot\lambda}{2fH}
	\,.\label{eq:lambda lambda_theta}
}
In the Einstein-de Sitter Universe with general relativity, $\kappa =\kappa_\theta =\lambda =\lambda_\theta =1$. In the case of the $\Lambda$CDM Universe, one shows $\kappa =\kappa_\theta =1$, but $\lambda$ and $\lambda_\theta$ deviate slightly from unity, with $\lambda\approx 1-\frac{3}{572}(1-\Omega_{\rm m})$~\cite{Bouchet:1994xp,Bernardeau:2001qr,Yamauchi:2017ibz}. If the gravity theory is described by the Horndeski scalar-tensor theories~\cite{Horndeski:1974wa,Deffayet:2011gz,Kobayashi:2011nu}, then $\kappa$ and $\kappa_\theta$ still take the standard values, but the time-dependence of $\lambda$ and $\lambda_\theta$ contains information from the underlying theory of gravity~\cite{Yamauchi:2017ibz,Takushima:2013foa}. In the DHOST scalar-tensor theories beyond Horndeski, 
not only $\lambda$ and $\lambda_\theta$, but also $\kappa$ and $\kappa_\theta$
can deviate from unity~\cite{Hirano:2018uar,Crisostomi:2019vhj,Lewandowski:2019txi}.

Hereafter, based on these variables let us discuss the degeneracy between the parameters by using the observation data from galaxy redshift surveys. 
To account for the uncertainty of the amplitude of the power spectrum, it is convenient to introduce $\sigma_8$ as the root-mean-square of the matter fluctuations averaged over $8\,h^{-1}{\rm Mpc}$. Using this parameter and Eq.~\eqref{eq:galaxy power spectrum}, one finds that the galaxy power spectrum can only constrain the combinations $Z_1({\bm k})\sigma_8$. In other words, the growth rate $f$ measured using the redshift-space distortion (RSD) cannot be independently determined and is always degenerate with $\sigma_8$ by galaxy power spectrum alone. As for the galaxy bispectrum, the situation is slightly changed because the shape dependence should be taken into account. To describe the shape dependence of the galaxy bispectrum induced by the quasi-nonlinear growth, we need to introduce the scale-dependent function related to the shift, which is defined as
\al{
	{\cal S}({\bm p}_1,{\bm p}_2)=\frac{1}{2}(\widehat{\bm p}_1\cdot\widehat{\bm p}_2)
			\left(\frac{p_1}{p_2}+\frac{p_2}{p_1}\right)
	\,,
}
in addition to the tidal term Eq.~\eqref{eq:tidal} and the scale-independent growth term. Moreover, since the galaxy bispectrum shown in Eq.~\eqref{eq:galaxy bispectrum} is in proportion to the square of the matter power spectrum, the $Z_2$ term is always measured by the combination with $\sigma_8^2$. When taking into account such observational effects, the measured $Z_2$ term from the galaxy bispectrum \eqref{eq:galaxy bispectrum} can be rewritten as
\al{
	Z_2({\bm k}_1,{\bm k}_2)\sigma_8^2
		=&(b_1\sigma_8)\sigma_8\biggl[
			\kappa{\cal S}({\bm k}_1,{\bm k}_2)
			+\left(\kappa -\frac{4}{21}\lambda -\frac{b_2}{2b_1}\right)
			+\left(\frac{2}{7}\lambda +\frac{b_{s^2}}{b_1}\right){\cal T}({\bm k}_1,{\bm k}_2)
		\biggr]
	\notag\\
	&\quad
		+(f\sigma_8 )(\widehat{\bm k}_3\cdot\widehat{\bm n})^2
			\sigma_8
			\biggl[
			\kappa_\theta {\cal S}({\bm k}_1,{\bm k}_2)
			+\left(\kappa_\theta -\frac{8}{21}\lambda_\theta \right)
			+\frac{4}{7}\lambda_\theta {\cal T}({\bm k}_1,{\bm k}_2)
			\biggr]
	\notag\\
	&\quad
			-\frac{(f\sigma_8)({\bm k}_3\cdot\widehat{\bm n})}{2}
				\Bigl[
					(\widehat{\bm k}_1\cdot\widehat{\bm n})Z_1({\bm k}_2)\sigma_8
					+(\widehat{\bm k}_2\cdot\widehat{\bm n})Z_1({\bm k}_1)\sigma_8
				\Bigr]
	\,.\label{eq:Z_2}
}
From this expression, we found that it is quite challenging to determine one of the coefficients of the second-order kernels, $\lambda$, from the measurement of the galaxy bispectrum, because $\lambda$ always appears with the nonlinear galaxy bias functions in the growth and the tidal terms, as shown in the first line of \eqref{eq:Z_2}. Namely, there is a strong degeneracy between $\lambda$, $b_2$, and $b_{s^2}$, and the signal of the gravity theory existing in the $\lambda$ term would be hidden by the uncertainty of the nonlinear galaxy bias. Moreover, since the shift term in the first line of \eqref{eq:Z_2} is measured by the combination of $\kappa\sigma_8$, $\kappa$ itself would not be suitable for extracting the information on the modified gravity. On the other hand, as for the term corresponding to the second-order peculiar velocity field, one finds there appear no bias contributions. Hence, we conclude that the second-order peculiar velocity field can be used to constrain the modification of gravity theory without suffering the uncertainty of the galaxy bias.

To avoid the degeneracy between $\sigma_8$ and other parameters related to the modified gravity theory, we propose new parametrizations:
\al{
	&E_{\rm f}=\frac{f\sigma_8}{\kappa\sigma_8}
	\,,\ \ \ 
	E_{\rm s}=\frac{\kappa_\theta\sigma_8}{\kappa\sigma_8}
	\,,\ \ \ 
	E_{\rm t}=\frac{\lambda_\theta\sigma_8}{\kappa\sigma_8}
	\,.\label{eq:E_s E_t def}
}
Using these parameters, we expect to easily extract meaningful information for gravity theory from the galaxy bispectrum without suffering from the parameter degeneracy of $\sigma_8$. 
Since $\kappa =\kappa_\theta =1$ in the case of general relativity and Horndeski scalar-tensor theories, $E_{\rm f}$ and $E_{\rm t}$ can be treated as $f$ and $\lambda_\theta$ inferred from observational data.
With these variables, \eqref{eq:Z_2} can be rewritten as
\al{
	Z_2({\bm k}_1,{\bm k}_2)\sigma_8^2
		=&(b_1\sigma_8)(f\sigma_8)E_{\rm f}^{-1}\biggl[
			{\cal S}({\bm k}_1,{\bm k}_2)
			+\left( 1-\frac{4\lambda}{21\kappa} -\frac{b_2}{2b_1\kappa}\right)
			+\left(\frac{2\lambda}{7\kappa} +\frac{b_{s^2}}{b_1\kappa}\right){\cal T}({\bm k}_1,{\bm k}_2)
		\biggr]
	\notag\\
	&\quad
		+(f\sigma_8 )^2E_{\rm f}^{-1}(\widehat{\bm k}_3\cdot\widehat{\bm n})^2
			\biggl[
			E_{\rm s}{\cal S}({\bm k}_1,{\bm k}_2)
			+\left( E_{\rm s} -\frac{8}{21}E_{\rm t} \right)
			+\frac{4}{7}E_{\rm t} {\cal T}({\bm k}_1,{\bm k}_2)
			\biggr]
	\notag\\
	&\quad
			-\frac{(f\sigma_8)({\bm k}_3\cdot\widehat{\bm n})}{2}
				\Bigl[
					(\widehat{\bm k}_1\cdot\widehat{\bm n})
					Z_1({\bm k}_2)\sigma_8
					+(\widehat{\bm k}_2\cdot\widehat{\bm n})
					Z_1({\bm k}_1)\sigma_8
				\Bigr]
	\,.\label{eq:Z2 exp}
}
This is one of the main results in this paper. Each term in Eq.~\eqref{eq:Z2 exp} can be observed independently 
by taking advantage of the different wave number dependence. Since the linear bias function $b_1\sigma_8$ and the linear growth rate $f\sigma_8$ are severely restricted by the observed galaxy power spectrum, we can use the galaxy bispectrum to extract the unbiased information of the parameters $E_{\rm f}$, $E_{\rm s}$, and $E_{\rm t}$ from the shift term in the first line, the shift-RSD and the tidal-RSD contributions in the second line of Eq.~\eqref{eq:Z2 exp}, respectively. As will be shown in the later sections, these three parameters allow us to trace the history of nonlinear growth and encompass the deviations within a broad theoretical framework. In the following section, we present theoretical predictions for the above parameters based on modified gravity theories, specifically the DHOST theories, and demonstrate the implications of these parameters for current and future galaxy redshift surveys.

\section{Nonlinear gravitational growth in degenerate higher-order scalar-tensor theories}
\label{sec:Nonlinear gravitational growth in degenerate higher-order scalar-tensor theories}

\subsection{Small-scale effective theory}

In order to describe the perturbations for metric and matter around a spatially flat FLRW solution, it is convenient to use the time-dependent parameters of effective-field-theory (EFT) of dark energy to specify the perturbations fully. For the DHOST theories, these for linear perturbations have been introduced in \cite{Langlois:2017mxy} and extended to nonlinear order in \cite{Dima:2017pwp}. In the context of the EFT, the metric is written in the ADM form:
\al{
	\dd s^2=-N^2\dd t^2+h_{ij}\left(\dd x^i+N^i\dd t\right)\left(\dd x^j+N^j\dd t\right)
	\,.
}
Choosing the time as to coincide with the uniform scalar-field hypersurface, as the perturbed variable, we consider the $\delta N=N-1$, the extrinsic curvature $\delta K^i{}_j=K^i{}_j-H\delta^i{}_j$ the three-dimensional spatial curvature ${}^{(3)}R$, with $H=\dot a/a$. With these variables, the effective Lagrangian is expressed as \cite{Langlois:2017mxy,Dima:2017pwp}
\al{
	{\cal L}_{\rm EFT}
		=&\sqrt{h}\frac{M^2}{2}
			\biggl[
				-\left( 1+\delta N\right)\delta{\cal K}_2
				+\left( 1+\alpha_{\rm T}\right){}^{(3)}R
				+H^2\alpha_{\rm K}\delta N^2
				+4H\alpha_{\rm B}\delta K\delta N
				+\left( 1+\alpha_{\rm H}\right){}^{(3)}R\delta N
	\notag\\
	&\quad\quad\quad\quad
				+4\beta_1\delta KV
				+\beta_2V^2
				+\beta_3a_ia^i
				+\alpha_{\rm V}\delta N\delta{\cal K}_2
			\biggr]
	\,,\label{eq:ADM effective Lagrangian}
}
where $\delta{\cal K}_2=\delta K^2-\delta K^i{}_j\delta K^j{}_i$,
$V=(\dot N-N^i\pd_i N)/N$, and $a_i=\pd_i N/N$.
In the EFT language, the degeneracy condition of the type-I DHOST theories 
to ensure the propagation of a single scalar degree of freedom 
reduces to
\al{
	\beta_2=-6\beta_1^2
	\,,\ \ \ 
	\beta_3=-2\beta_1\Bigl[2\left(1+\alpha_{\rm H}\right)+\beta_1\left(1+\alpha_{\rm T}\right)\Bigr]
	\,.
}
For later convenience, we introduce another time-dependent function:
\al{
	&\alpha_{\rm M}:=\frac{\dd\ln (M^2)}{\dd\ln a}
	\,.
}
The minimum set to fully specify the total amount of cosmological perturbations up to the linear order in the type-I DHOST theories is six independent functions of time that are labelled $\alpha_{\rm T}$, $\alpha_{\rm K}$, $\alpha_{\rm B}$, $\alpha_{\rm M}$, $\alpha_{\rm H}$, and $\beta_1$ in addition to the Hubble parameter $H$ and the effective Planck mass $M$. To take into account the second-order perturbations, we need to consider the additional time-dependent function $\alpha_{\rm V}$, which is originally introduced in \cite{Yamauchi:2017ibz,Bellini:2015wfa} in the context of the Horndeski scalar-tensor theories.

In order to study cosmological perturbations, it is convenient to change the gauge to compare the standard results. To do so, we need to recover the scalar degree of freedom. In this section, we perform a time coordinate transformation $t\to t+\pi(t,{\bm x})$ and consider the Newtonian gauge given by
\al{
	\dd s^2
		=-\Bigl[1+2\Phi(t,{\bm x})\Bigr]\dd t^2
			+a^2(t)\Bigl[1-2\Psi (t,{\bm x})\Bigr]\delta_{ij}\dd x^i\dd x^j
	\,.
}
We then introduce a dimensionless variable $Q(t,{\bm x})=H\pi (t,{\bm x})$.
The nonrelativistic matter energy is given by
\al{
	\rho(t,{\bm x})=\rho_{\rm m}(t)\Bigl[ 1+\delta (t,{\bm x})\Bigr]
	\,.
}
To study the quasi-static behaviour deep inside the horizon, we expand the action in terms of the metric and the scalar field perturbations \cite{Dima:2017pwp,Kobayashi:2014ida,Hirano:2019scf,Hiramatsu:2020fcd}. In the quasi-static regime, the time derivatives of those perturbations are of order Hubble and much smaller than their spatial derivatives. Moreover, the Lagrangian is dominated by terms with $2(n+1)$ spatial derivatives for $n+2$ fields. Namely, we will keep the terms of the form of $(\pd\epsilon)^2(\pd^2\epsilon)^n$ in the action, where $\epsilon$ stands for any of $\Phi$, $\Psi$, $Q$ and their time derivatives. The matter overdensity $\delta$ is assumed to be of ${\cal O}(\pd^2\epsilon )$. We note that we should keep the mixed derivative terms such as $\pd^2\dot Q/H$, which cannot be simply ignored, as shown in \cite{Kobayashi:2014ida}. By expanding the action following the above rule, we obtain the small-scale effective Lagrangian of the form
\al{
	{\cal L}_{\rm EFT}={\cal L}_2+{\cal L}_3+\cdots
	\,,
}
where ${\cal L}_n$ denote the $n$-th order terms, which are explicitly given by
\al{
	{\cal L}_2
		=&\frac{M^2a}{2}
			\biggl[
				4\left(1+\alpha_{\rm H}\right)\Psi\pd^2\Phi
				-2\left( 1+\alpha_{\rm T}\right)\Psi\pd^2\Psi
				-\beta_3\Phi\pd^2\Phi
	\notag\\
	&\quad\quad
	       +4\biggl\{\alpha_{\rm M}-\alpha_{\rm T}+\frac{(aM^2\alpha_{\rm H})^\cdot}{aM^2H}\biggr\}\Psi\pd^2 Q
	       -4\biggl\{\alpha_{\rm B}-\alpha_{\rm H}+\frac{(aM^2\beta_3)^\cdot}{2aM^2H}\biggr\}\Phi\pd^2 Q
				+c_{QQ}Q\pd^2 Q
	\notag\\
	&\quad\quad
			+\biggl\{ 4\alpha_{\rm H}\frac{\dot\Psi}{H}-2\left(2\beta_1+\beta_3\right)\frac{\dot\Phi}{H}+\left( 4\beta_1+\beta_3\right)\frac{\ddot Q}{H^2}\biggr\}\pd^2 Q
			\biggr]-a^3\rho_{\rm m}\Phi\delta
	\,,\label{eq:L2}
}
and
\al{
	{\cal L}_3
		=&\frac{M^2}{2aH^2}
			\biggl[
				-\frac{1}{2}c_{QQQ}(\pd Q)^2\pd^2 Q
				+\Bigl\{
					\left(\alpha_{\rm V}-\alpha_{\rm H}-4\beta_1\right)\Phi
					+\alpha_{\rm T}\Psi
				\Bigr\}\Bigl\{(\pd^2 Q)^2-(\pd_i\pd_j Q)^2\Bigr\}
	\notag\\
	&\quad\quad
				+\biggl\{
					-4\alpha_{\rm H}\pd_i\Psi
					+2\left( 2\beta_1+\beta_3\right)\pd_i\Phi
					-2\left( 4\beta_1+\beta_3\right)\frac{\pd_i\dot Q}{H}
				\biggr\}\pd_j Q\pd_i\pd_j Q
			\biggr]
	\,.\label{eq:L3}
}
Here, the explicit forms of $c_{QQ}$ and $c_{QQQ}$ 
are given 
\al{
    c_{QQ} &= 
        -2\biggl\{\,
		\frac{\dot H}{H^2}+\frac{3}{2}\Omega_{\rm m}
		+\alpha_{\rm T}-\alpha_{\rm M}
		+\frac{[aM^2H(\alpha_{\rm B}-\alpha_{\rm H})]^\cdot}{aM^2H^2}
		+\frac{(aM^2\beta_3)^{\cdot\cdot}}{4aM^2H^2}
		-\frac{\dot H}{2H^2}\left(\frac{aM^2(4\beta_1+\beta_3 )}{H}\right)^\cdot\,
	\biggr\}
    \,,\\
    c_{QQQ} &=
    -\left[\alpha_{\rm V} +3(\alpha_{\rm H} -\alpha_{\rm T}) -4\alpha_{\rm B}
        +\alpha_{\rm M}(2 -\alpha_{\rm V} +\alpha_{\rm H} +8\beta_{1})
        +2(4\beta_{1} +\beta_{3})\frac{\dot H}{H^{2}}
        -\frac{\dot \alpha_{\rm V} -\dot\alpha_{\rm H} -8\dot\beta_{1}}{H}\right]
    \,,
}
with $\Omega_{\rm m}=\rho_{\rm m}/3M^2H^2$.
The coefficients and the time-dependent functions related to the scalar-tensor theories and should be evaluated on the background.

\subsection{Evolution equation for density fluctuations}

By varying the Lagrangian with respect to $\Phi$, $\Psi$, and $Q$, 
and solving them in terms of $\delta$ and its time derivatives, 
we can formally write the effective Poisson equation valid up to the second-order \cite{Takushima:2013foa,Takushima:2015iha,Hirano:2020dom}
\al{
	-\frac{k^2}{a^2H^2}&\Phi (t,{\bm k})
		=\kappa_\Phi (t)\delta (t,{\bm k}) +\nu_\Phi (t)\frac{\dot{\delta} (t,{\bm k})}{H}
			+\mu_\Phi (t)\frac{\ddot{\delta} (t,{\bm k})}{H^2}
	\notag\\
	&
			+\int\frac{\dd^3{\bm p}_1\dd^3{\bm p}_2}{(2\pi)^3}
				\delta_{\rm D}^3({\bm k}-{\bm p}_1-{\bm p}_2)
				\Bigl[\tau_{\Phi ,\alpha} (t)\alpha_{\rm s}({\bm p}_1,{\bm p}_2)+\tau_{\Phi,\gamma}(t)\gamma ({\bm p}_1,{\bm p}_2)\Bigr]\delta(t,{\bm p}_1)\delta(t,{\bm p}_2)
			+\cdots
	\,,\label{eq:effective Poisson}
}
where the second-order mode coupling functions
$\alpha_{\rm s}({\bm p}_1,{\bm p}_2)$ and $\gamma ({\bm p}_1,{\bm p}_2)$
were defined in Eqs.~\eqref{eq:mode coupling alpha_s def} and
\eqref{eq:mode coupling gamma def}.
Here, the time-dependent functions $\kappa_\Phi$, $\nu_\Phi$, $\mu_\Phi$, $\tau_{\Phi,\alpha}$ and $\tau_{\Phi,\gamma}$ are related to the EFT parameters $\alpha_i$ ($i={\rm T},{\rm B},{\rm M},{\rm H},{\rm V}$) and $\beta_{1,2,3}$ 
appearing in Eq.~\eqref{eq:ADM effective Lagrangian}.
Their relations are shown in Appendix \ref{sec:Coefficients of first- and second-order solutions}.

Throughout this paper, we assume that the matter is minimally coupled to gravity. 
The continuity and Euler equations for the pressureless nonrelativistic matter are given by
\al{
	&\dot\delta +\frac{1}{a}\pd_i\bigl[(1+\delta )v^i\bigr] =0
	\,,\\
	&\dot v_i+Hv_i+\frac{1}{a}v^j\pd_j v^i=-\frac{1}{a}\pd^i\Phi
	\,.
}
Although these fluid equation are same as the standard ones in general relativity, there appears the effect of the modification of gravity theory through the effective Poisson equation \eqref{eq:effective Poisson}.
Combining these with Eq.~\eqref{eq:effective Poisson} to eliminate the gravitational potential $\Phi$, 
we derive the closed-form equation for the linear growth of the density fluctuation, $D_+(t)$, in the form
\al{
	\ddot D_++\left( 2+\varsigma\right) H\dot D_+-\frac{3}{2}\Omega_{\rm m}\Xi H^2D_+ =0
	\,,\label{eq: D eq}
}
where $\varsigma :=(2\mu_\Phi -\nu_\Phi )/(1-\mu_\Phi )$, $(3/2)\Omega_{\rm m}\Xi :=\kappa_\Phi /(1-\mu_\Phi )$. 
Once the time-dependent coefficients $\varsigma$ and $\Xi$ are given, one can solve this equation
with the boundary condition given by $D_+\propto a$ at $a\ll 1$.
This equation can be reinterpreted as the evolution equation for the linear growth rate $f=\dd\ln D_+/\dd\ln a$ as
\al{
	\frac{\dd f}{\dd\ln a}
		+\left( 2+\varsigma +\frac{\dd\ln H}{\dd\ln a}\right) f+f^2-\frac{3}{2}\Omega_{\rm m}\Xi =0
	\,.\label{eq:f eq}
}
This means that the precise measurement of the linear growth rate from the redshift space distortion can provide the information captured
in $\varsigma$ and $\Xi$.
To investigate the second-order nonlinear growth of structure, we need to take into account the nonlinear mode coupling terms as the source term.
The equation to solve is written as
\al{
	&\ddot\delta (t,{\bm k})
		+\left( 2+\varsigma \right) H\dot\delta (t,{\bm k}) -\frac{3}{2}\Omega_{\rm m}\Xi H^2\delta (t,{\bm k}) 
    \notag\\
    &\quad\quad
    =H^2D_+^2\int\frac{\dd^3{\bm p}_1\dd^3{\bm p}_2}{(2\pi)^3}\delta_{\rm D}^3({\bm k}-{\bm p}_1-{\bm p}_2)
            \Bigl[
                S_{\kappa} (t)\alpha_{\rm s}({\bm p}_1,{\bm p}_2)+S_\lambda (t)\gamma ({\bm p}_1,{\bm p}_2)
            \Bigr]\delta_{\rm L}({\bm p}_1)\delta_{\rm L}({\bm p}_2)
	\,,\label{eq:2nd order delta eq}
}
where the nonlinear coefficients are given by \cite{Hirano:2020dom}
\al{
	&S_\kappa =\frac{1}{1-\mu_\Phi}\left( 2f^2+\frac{3}{2}\Omega_{\rm m}\Xi -\varsigma f+\tau_{\Phi,\alpha}\right)
	\,,\\
	&S_\lambda =\frac{7}{2(1-\mu_\Phi )}\left( -f^2+\tau_{\Phi,\gamma}\right)
	\,.
}
Reminding Eqs.~\eqref{eq:delta expansion} and \eqref{eq:F_2}, we can rewrite Eq.~\eqref{eq:2nd order delta eq}
as the equation of $\zeta =\kappa\,,\lambda$:
\al{
	&\frac{\dd^2\zeta}{\dd\ln a^2}+\left( 2+\varsigma+\frac{\dd\ln H}{\dd\ln a}+4f\right)\frac{\dd\zeta}{\dd\ln a}
		+\left( 2f^2+\frac{3}{2}\Omega_{\rm m}\Xi\right)\zeta
		=S_\zeta
	\,.\label{eq:zeta eq}
}
The second-order kernels $F_2$ and $G_2$ should coincide with the well-known results in the Einstein-de Sitter Universe in the deep matter-dominated era. Hence, we impose the boundary conditions: $\kappa =1$ and $\lambda =1$ at $a\ll 1$. This shows that the time-dependent coefficients in the second-order $F_2$ kernel, $\kappa$ and $\lambda$, can carry the information not only about $\varsigma$ and $\Xi$ but also about the nonlinear interaction terms $\tau_{\Phi,\alpha}$ and $\tau_{\Phi,\gamma}$.
Given the solutions of $\kappa$ and $\lambda$ by solving the above evolution equation, we can obtain the second-order coefficients of the peculiar velocity field $\kappa_\theta$
and $\lambda_\theta$ from Eqs.~\eqref{eq:kappa kappa_theta} and \eqref{eq:lambda lambda_theta}.

\section{Approximate expression}
\label{sec:Approximate expression}

\subsection{Setup}
\label{sec:Setup}

In this section, we consider the approximate expression of the nonlinear growth functions $\kappa$, $\lambda$, $\kappa_\theta$, and $\lambda_\theta$ in order to derive the analytic formulae of the second-order variables $E_{\rm f}$, $E_{\rm s}$, and $E_{\rm t}$ [Eq.~\eqref{eq:E_s E_t def}] in addition to the linear growth rate. To do so, we need to specify the background expansion history of the Universe. First, we write down the Friedmann and the matter conservation equation in terms of $\Omega_{\rm m}$ as
\al{
	& \frac{\rho_{\rm DE}}{3M^2H^2}=1-\Omega_{\rm m}
	\,,\\
	&\frac{\dd\ln H}{\dd\ln a}=-\frac{3}{2}\Bigl[ 1+w\left( 1-\Omega_{\rm m}\right)\Bigr]
	\,,\label{eq:dot H}\\
	&\frac{\dd\ln\Omega_{\rm m}}{\dd\ln a}=3w\left( 1-\Omega_{\rm m}\right) -\alpha_{\rm M}
	\,,\label{eq:dotOmega_m}
}
where $w$ denotes a dark-energy effective equation-of-state parameter $w=P_{\rm DE}/\rho_{\rm DE}$ for the dark energy component, whose energy density and pressure are defined in terms of the Hubble parameter and effective Planck mass through $\rho_{\rm DE}:= 3M^2H^2-\rho_{\rm m}$, $P_{\rm DE}:= -M^2(3H^2+2\dot H)$.

In order to solve Eqs.~\eqref{eq:f eq} and \eqref{eq:zeta eq} analytically, 
we assume that the Universe can be well described by the $\Lambda$CDM model 
and the excitation of the scalar field is sufficiently suppressed in the deep matter dominated era, 
and we focus only on the matter dominated era and the early stage of the dark energy dominated era.
During the era of interest, we can treat $\varepsilon =1-\Omega_{\rm m}$ 
as a expansion parameter~\footnote{
In this treatment, we treat $\Omega_{\rm m}$ as a time variable.
However, when one applies our formalism to observational data, 
the redshift-dependence of $\Omega_{\rm m}$ would be needed.
Our definition of $\Omega_{\rm m}$ depends on not only the background evolution but also
the effective Planck mass running, as seen in Eq.~\eqref{eq:dotOmega_m}.
The difference between the standard $\Omega_{\rm m}$ defined in the $\Lambda$CDM Universe with
general relativity and ours is discussed in Appendix \ref{sec:Diffrent parametrzation}.
}.
Hence, the equation-of-state parameter $w$, the EFT parameters $\alpha_i$ 
($i={\rm T},{\rm B},{\rm M},{\rm H},{\rm V}$) and $\beta_1$ can be expanded 
as a series expansion form
in terms of $\varepsilon$ as
\al{
	&w=w^{(0)}+{\cal O}\left(\varepsilon \right)
	\,,\label{eq:w exp}\\
	&\alpha_i =c_i\varepsilon +{\cal O}(\varepsilon^2)
	\,,\\
	&\beta_1 =\beta\varepsilon +{\cal O}(\varepsilon^2)
	\,,\label{eq:beta_1 exp}
}
where $w^{(0)}$, $c_i$, and $\beta$ are constant parameters,
which should be evaluated at the deep matter dominated era.
Since the equation-of-motions for the linear- and second-order growth of the density fluctuation Eqs.~\eqref{eq: D eq} and \eqref{eq:2nd order delta eq} at the deep matter dominated era is also assumed to be consistent with the standard one in the $\Lambda$CDM model, their deviation should be suppressed by the factor $\varepsilon$. Therefore,
it is expected that the early-time asymptotes of $\mu_\Phi$, $\varsigma$, $\Xi$, and $\tau_{\Phi,\Pi}$  for $\varepsilon\to 0$ are $\mu_\Phi\to 0$, $\varsigma\to 0$, $\Xi\to 1$, and $\tau_{\Phi,\Pi}\to 0$. These variables during the era of interest can be written as
\al{
	&\mu_\Phi =\mu_\Phi^{(1)}\varepsilon +{\cal O}(\varepsilon^2)
	\,,\label{eq:mu exp}\\
	&\varsigma =\varsigma^{(1)}\varepsilon +{\cal O}(\varepsilon^2)
	\,,\\
	&\Xi =1+\Xi^{(1)}\varepsilon +{\cal O}(\varepsilon^2)
	\,,\label{eq:Xi exp}\\
	&\tau_{\Phi,\Pi}=\tau_{\Phi,\Pi}^{(1)}\varepsilon +{\cal O}(\varepsilon^2 )
	\,.\label{eq:tau_Pi exp}
}
Here, the first-order coefficients $\mu_\Phi^{(1)}$, $\varsigma^{(1)}$, $\Xi^{(1)}$, and $\tau_{\Phi,\Pi}^{(1)}$ 
can be written in terms of the expansion parameters defined in Eqs.~\eqref{eq:w exp}--\eqref{eq:beta_1 exp}.
The explicit form of these are presented in Appendix \ref{sec:Explicit expression for some coefficients}.

\subsection{First-order}

Let us first solve the first order equation \eqref{eq:f eq} to express the growth rate $f$, following Ref.~\cite{Hirano:2019nkz}.
Based on the assumption described in the previous subsection, the equation for the growth rate $f$ 
can reduce to
\al{
    \left( c_{\rm M}-3w^{(0)}\right)\varepsilon\frac{\dd f}{\dd\varepsilon}
    +\biggl[\frac{1}{2}+\left(\varsigma^{(1)}-\frac{3}{2}w^{(0)}\right)\varepsilon\biggr] f
    +f^2-\frac{3}{2}\biggl[1-\left(1-\Xi^{(1)}\right)\varepsilon\biggr]
    +{\cal O}(\varepsilon^2)
    =0
    \,.
}
Since the growth rate $f$ approaches to unity at the deep matter dominated era,
we can solve the above equation to obtain the next-leading order solution of $f$ in terms of $\varepsilon$ as
\al{
	f=1-\biggl[\frac{3(1-w^{(0)})+2\varsigma^{(1)}-3\Xi^{(1)}}{5-6w^{(0)}+2c_{\rm M}}\biggr]\varepsilon 
		+{\cal O}(\varepsilon^2)
	\,,
}
which immediately implies that the corresponding growth index $\gamma$, which was defined in Eq.~\eqref{eq:growth index def}, is 
given by
\al{
	\gamma =\frac{3(1-w^{(0)})+2\varsigma^{(1)}-3\Xi^{(1)}}{5-6w^{(0)}+2c_{\rm M}} +{\cal O}(\epsilon )
	\,.\label{eq:gamma sol}
}
Substituting the explicit expressions of 
$\varsigma^{(1)}$ [Eq.~\eqref{eq:varsigma1}] and $\Xi^{(1)}$ [Eq.~\eqref{eq:Xi1}] 
into Eq.~\eqref{eq:gamma sol}, we obtain
\al{
	\gamma =&\frac{3[(1-w^{(0)})-c_{\rm T}+2(c_{\rm H}+\beta)]}{5-6w^{(0)}+2c_{\rm M}}
			-\frac{2\rho^2}{\Sigma}
	\notag\\
	&
	        -\frac{c_{\rm H}+\beta}{\Sigma}
					\biggl\{
						\left( 3+2c_{\rm M}-6w^{(0)}\right)\rho
						+\left( 2-c_{\rm M}+3w^{(0)}\right)\left( c_{\rm H}+\beta\right)
					\biggr\}
			+{\cal O}(\varepsilon)
	\,,\label{eq:gamma exp}
}
where
\al{
    &\rho =c_{\rm B}-c_{\rm M}+c_{\rm T}-\beta\left( c_{\rm M}-3w^{(0)}\right)
    \,,\\
	&\Sigma =\frac{1}{6}\left( 5-6w^{(0)}+2c_{\rm M}\right)Z^{(1)}
	\,.
}
with
\al{
	&Z^{(1)}=2\bigg\{
			3\left( 1+w^{(0)}\right) +2\left( c_{\rm M}-c_{\rm T}\right) 
			+\Bigl[ 1-2\left(c_{\rm M}-3w^{(0)}\right)\Bigr]\Bigl[ c_{\rm B}-c_{\rm H}-\beta\left( c_{\rm M}-3w^{(0)}+1\right)\Bigr]
		\biggr\}
	\,.\label{eq:Z1 def}
}

\subsection{Second-order}

We next derive the solution of the second-order equation-of-motion for the density fluctuation under the assumptions discussed in Sec.~\ref{sec:Setup}. We can formally solve the equations for the second-order coefficients $\kappa$ and $\lambda$, Eq.~\eqref{eq:zeta eq}. The corresponding solutions are then written as
\al{
	&\kappa =1-\biggl[\frac{-2\tau_{\Phi,\alpha}^{(1)}+2\varsigma^{(1)}-7\mu_\Phi^{(1)}}{(7-6w^{(0)}+2c_{\rm M})(1-3w^{(0)}+c_{\rm M})}\biggr]\varepsilon +{\cal O}(\varepsilon^2)
	\,,\\
	&\lambda =1-\biggl[\frac{-3+6\gamma -7\tau_{\Phi,\gamma}^{(1)}+3\Xi^{(1)}-7\mu_\Phi^{(1)}}{(7-6w^{(0)}+2c_{\rm M})(1-3w^{(0)}+c_{\rm M})}\biggr]\varepsilon +{\cal O}(\varepsilon^2)
	\,.
}
The corresponding second-order indexes Eq.~\eqref{eq:second-order index def} are expressed as
\al{
	&\xi_\kappa =\frac{-2\tau_{\Phi,\alpha}^{(1)}+2\varsigma^{(1)}-7\mu_\Phi^{(1)}}{(7-6w^{(0)}+2c_{\rm M})(1-3w^{(0)}+c_{\rm M})}
			+{\cal O}(\varepsilon )
	\,,\label{eq:xi_kappa sol}\\
	&\xi_\lambda =\frac{-3+6\gamma-7\tau_{\Phi,\gamma}^{(1)}+3\Xi^{(1)}-7\mu_\Phi^{(1)}}{(7-6w^{(0)}+2c_{\rm M})(1-3w^{(0)}+c_{\rm M})}
			+{\cal O}(\varepsilon )
	\,.\label{eq:xi_lambda sol}
}
With the use of the explicit expression of $\mu_\Phi^{(1)}$ and $\tau_{\Phi,\Pi}^{(1)}$ given in
Eqs.~\eqref{eq:mu1} and \eqref{eq:tau_Phi,alpha}, in addition to $\mu^{(1)}$, $\varsigma^{(1)}$, and $\Xi^{(1)}$,
we obtain the approximate solutions of the second-order index $\xi_\kappa$ and $\xi_\lambda$
during the matter dominated era and the early stage of the dark energy dominated era.
We first show the explicit expression of $\xi_\kappa$ as
\al{
	\xi_\kappa
		=\frac{6(c_{\rm H}+\beta )^2(1-K_Q^{(0)})}{(1+c_{\rm M}-3w^{(0)})Z^{(1)}}+{\cal O}(\varepsilon)
	\,,\label{eq:xi_kappa exp}
}
where $Z^{(1)}$ was defined in Eq.~\eqref{eq:Z1 def}, and 
$K_Q^{(0)}$ denotes the leading order solution of the scalar-field perturbation 
(see Eqs.~\eqref{eq:K def} and \eqref{eq:K exp}), which is explicitly given by
\al{
    K_Q^{(0)} =&\frac{6\Bigl[\rho-\left(c_{\rm H}+\beta\right)\left( 2+c_{\rm M}-3w^{(0)}\right)\Bigr]}{Z^{(1)}}
    \,.
}
The new parameters $E_{\rm f}$ and $E_{\rm s}$ defined in Eq.~\eqref{eq:E_s E_t def} can be rewritten in terms
of $\gamma$ and $\xi_\kappa$ as
\al{
    &E_{\rm f}
        =1+\left(\xi_\kappa -\gamma\right)\varepsilon +{\cal O}(\varepsilon^2)
    \,,\label{eq:E_f exp}\\
    &E_{\rm s}
        =1-\left(1+c_{\rm M}-3w^{(0)}\right)\xi_\kappa\,\varepsilon +{\cal O}(\varepsilon^2)
    \,,
}
which immediately lead to
\al{
    &\xi_{\rm f}
        :=\frac{\dd\ln E_{\rm f}}{\dd\ln\Omega_{\rm m}}
        =\gamma -\frac{6(c_{\rm H}+\beta )^2(1-K_Q^{(0)})}{(1+c_{\rm M}-3w^{(0)})Z^{(1)}}+{\cal O}(\varepsilon)
    \,,\\
	&\xi_{\rm s}
	    :=\frac{\dd\ln E_{\rm s}}{\dd\ln\Omega_{\rm m}}
	    =\frac{6( c_{\rm H}+\beta )^2(1-K_Q^{(0)})}{Z^{(1)}} +{\cal O}(\varepsilon)
	\,.\label{eq:xi_s}
}
We found that the nontrivial time-dependence can capture the information of the modification 
of the gravity theory.
An important observation is that in the case of the general relativity 
and the Horndeski scalar-tensor theories,
that is $c_{\rm H}=\beta =0$, one can easily show $\xi_{\rm f}=\gamma$ and $\xi_{\rm s}=0$.
Therefore, we conclude that any nonvanishing value of $\xi_{\rm s}$
can be treated as the clear signal of the existence of the gravity theory 
beyond the Horndeski scalar-tensor theories.

Next, let us evaluate $\xi_\lambda$ in terms of the EFT parameters. 
We rewrite Eq.~\eqref{eq:xi_lambda sol} by using Eqs.~\eqref{eq:gamma sol}, \eqref{eq:mu1}, and \eqref{eq:varsigma1}:
\al{
    \xi_\lambda =&\frac{(1-2c_{\rm M}+6w^{(0)})\gamma -3w^{(0)} -7\tau_{\Phi,\gamma}^{(1)}}{(7+2c_{\rm M}-6w^{(0)})(1+c_{\rm M}-3w^{(0)})}
			+\frac{6(c_{\rm H}+\beta)^2}{(1+c_{\rm M}-3w^{(0)})Z^{(1)}}
			+{\cal O}(\varepsilon)
	\,.\label{eq:xi_lambda exp}
}
The leading term of the nonlinear mode-coupling, $\tau_{\Phi,\gamma}^{(1)}$, is given by
\al{
    \tau_{\Phi ,\gamma}^{(1)}
		=&-\frac{1}{4}\left(K_Q^{(0)}\right)^2
		    \left(c_{\rm T}+c_{\rm V}-5c_{\rm H}-12\beta\right)
		    -\frac{K_Q^{(0)}}{Z^{(1)}}
		    \biggl[\rho+\frac{3}{2}(c_{\rm H}+\beta)\biggr]
		    \biggl[ 
	            3\left( c_{\rm T}+c_{\rm V}-c_{\rm H}-4\beta\right)
	        +c_{QQQ}^{(1)}K_Q^{(0)}
	        \biggr]
    \,,\label{eq:tau_gamma1}
}
with
\al{
    c_{QQQ}^{(1)}
            =4c_{\rm B}-2c_{\rm M}+3c_{\rm T}-3c_{\rm H}-c_{\rm V}+\left( c_{\rm V}-c_{\rm H}-8\beta\right)\left( c_{\rm M}-3w^{(0)}\right)
    \,.
}
We then translate the second-order index $\xi_\kappa$ and $\xi_\lambda$
derived here into the new variable $E_{\rm t}$ [Eq.~\eqref{eq:E_s E_t def}]:
\al{
	E_{\rm t}
		 =1+\biggl[
		    \xi_\kappa -\frac{1}{2}\left(2+c_{\rm M}-3w^{(0)}\right)\xi_\lambda
		   \biggr]\varepsilon +{\cal O}(\varepsilon^2)
	\,.\label{eq:E_t exp}
}
Substituting Eqs.~\eqref{eq:xi_kappa exp}, \eqref{eq:xi_lambda exp}, and \eqref{eq:tau_gamma1} into Eq.~\eqref{eq:E_t exp},
$E_{\rm t}$ can be obtained in terms of constant parameters $\{w^{(0)},c_{\rm B},c_{\rm M},c_{\rm T},c_{\rm H},\beta\}$ and $c_{\rm V}$, but its explicit expression is complicated. Moreover, the corresponding index $\xi_t$ can be defined in the same way as $E_{\rm s}$ through the logarithmic derivative of $E_{\rm t}$ with respect to $\Omega_{\rm m}$. Given the $\Lambda$CDM Universe with general relativity, i.e. $w^{(0)}=-1$ and $c_i=\beta =0$, we can reproduce the standard result $\xi_\lambda =\frac{3}{572}$~\cite{Bouchet:1994xp,Bernardeau:2001qr}, which corresponds to $\xi_{\rm t}=\frac{15}{1144}$. Unlike $\xi_{\rm s}$, 
$\xi_{\rm t}$ can deviate from the standard value even when $c_{\rm H}=\beta =0$. Therefore, we can use the time-dependence of $E_{\rm t}$ to constrain the Horndeski scalar-tensor theories.

\subsection{Constraining DHOST cosmology with growth and
second-order peculiar velocity field}
\label{sec:Constraining DHOST cosmology}

In this subsection, as an application, we apply the resultant formula derived in the previous subsection to the shift-symmetric DHOST model developed in Ref.~\cite{Hirano:2019nkz}. 
This model allows us to consistently solve the equations for the evolution of both the background and perturbation during the matter-dominated era and the early stage of dark energy-dominated era, and to rewrite the EFT parameters that characterize the DHOST theories in terms of four constant parameters. To proceed with analysis, we focus on the specific type of the DHOST Lagrangian. We assume that the k-essence term in the DHOST theories is in proportion to $X^p$, 
where $p$ is a constant model parameter. Furthermore, we consider a tracker solution whose scalar field satisfies the condition $H\dot\phi^{2q}={\rm const}$ with $q$ being another constant parameter. For instance, in the DHOST cosmological model proposed in \cite{Crisostomi:2017pjs}, there is a cosmological solution that exhibits the late-time self-acceleration regime, corresponding to the case $p=1$ and $q=1/2$. Under these assumptions and considering that the cosmological solution is well described by an attractor solution, the leading order coefficients of the equation-of-state parameter $w^{(0)}$ and the braiding parameter $c_{\rm B}$ are shown to be written in terms of $p$, $q$, $c_{\rm H}$, and $\beta$ as
\al{
	&w^{(0)}=-1+\frac{c_{\rm H}+2p}{4q}
	\,,\\
	&c_{\rm B}=-p-\frac{c_{\rm H}}{3}-\frac{3\beta (2q-1)}{4q}
	\,.
}
We further consider that the gravitational wave event GW170817~\cite{TheLIGOScientific:2017qsa} and its optical counterpart GRB170817A~\cite{Monitor:2017mdv} were detected almost simultaneously, providing the stringent constraint on the deviation of the speed of gravitational waves from that of light. This measurement strongly implies that the speed of gravitational waves is in exact agreement with the speed of light. Even with this condition imposed, a certain subclass of type-I DHOST theories survived~\cite{Langlois:2017dyl,Creminelli:2017sry}. The gravitational waves travel at the same speed
as light, unaffected by slight changes in the background, when~\cite{Dima:2017pwp}
\al{
    \alpha_{\rm T}=0
    \,,\ \ \ 
    \alpha_{\rm V}=-\alpha_{\rm H}
    \,,
}
which corresponds to
\al{
	c_{\rm T}=0
	\,,\ \ 
	c_{\rm V}=-c_{\rm H}
	\,.
}
When imposing the above conditions,
the explicit forms of $\alpha_{\rm M}$
and $\alpha_{\rm H}$ in this setup 
implies the additional relation:
\al{
	c_{\rm M}=\frac{3}{4q}c_{\rm H}
	\,.
}
Combining these relations, we finally have independent parameters 
$(p,q,c_{\rm H},\beta)$
to model the shift-symmetric DHOST cosmology during the matter
dominated era and the early stage of the dark energy dominated 
era.~\footnote{Recently, the constraint from the stability
of gravitons against decay into dark energy and the gradient instability 
induced by gravitational waves are discussed in the literature~\cite{Creminelli:2018xsv,Creminelli:2019kjy}.
However, the case with these conditions is shown to be
the special class, in which for instance the screening mechanism
works only when the parameter fine-tuning is imposed~\cite{Hirano:2019scf,Crisostomi:2019yfo}.
Hence, it is beyond the scope of this paper and we simply
neglect these possibilities.
See also \cite{Bahamonde:2019ipm,Bahamonde:2019shr} for other possibilities.
}
The growth index $\gamma$, the second-order growth index $\xi_{\rm s}$
can be written in terms of $(p,q,c_{\rm H},\beta )$ as
\al{
	&\gamma
		=\frac{3}{2(-3+6p+10q)(3p+11q)}
			\biggl\{
				\Bigl[\left( p+4q\right)\left( -3+6p+10q\right) -8pq^2\Bigr]
	\notag\\
	&\quad\quad\quad
				+\frac{1}{2}\Bigl[ \left( -3+6p+10q\right)+8q\left(3p+2q\right)\Bigr] c_{\rm H}
				+\frac{3q(1+2q)(-3+6p+16q)(c_{\rm H}+\beta)^2}{2pq+3qc_{\rm H}+(3p+5q)\beta}
			\biggr\}+{\cal O}(\varepsilon)
	\,,\label{eq:gamma pq}\\
	&\xi_{\rm s} =\frac{12(3-6p-16q)(2pq+6qc_{\rm H}+(3p+8q)\beta )(c_{\rm H}+\beta)^2}{(3-6p-10q)^2(2pq+3qc_{\rm H}+(3p+5q)\beta)^2} +{\cal O}(\varepsilon)
	\,.\label{eq:xis pq}
}
We can also express $\xi_{\rm t}$ in terms of the four constant parameters 
$(p,q,c_{\rm H},\beta )$,
while its explicit form is not shown here.

\begin{figure}
    \includegraphics[height=80mm]{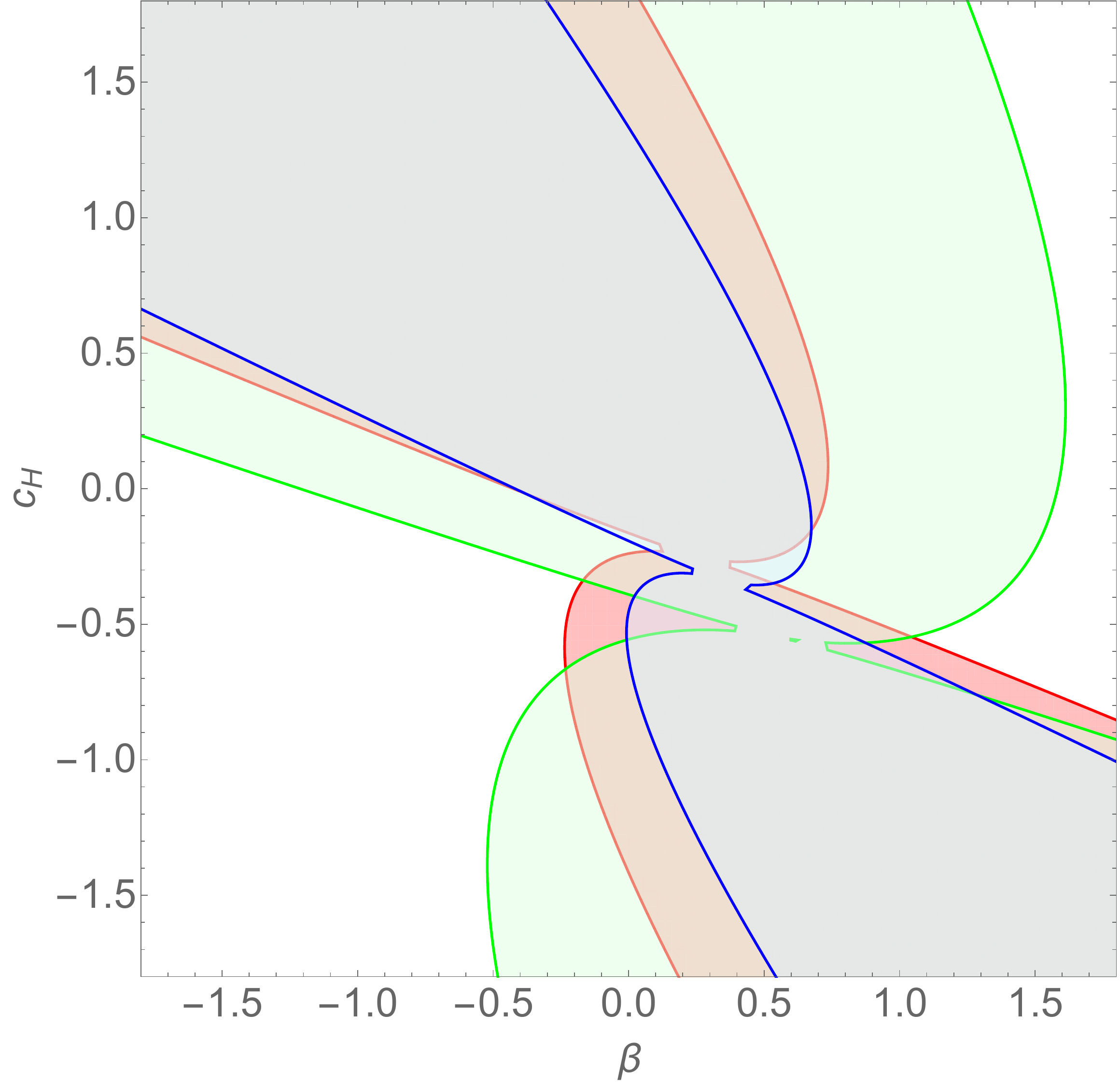}
     \caption{
     The allowed parameter region in the $(\beta ,c_{\rm H})$ plane obtained
     from the condition $|\xi_{\rm s}|\leq 0.2$ in the shift-symmetric DHOST cosmology. 
     The parameters are given by $(p,q)=(1,1/2)$ (red), $(1,3/4)$ (blue), and $(3,1)$ (green).
	}
     \label{fig:plot_xis}
\end{figure}

\begin{figure}
    \includegraphics[height=80mm]{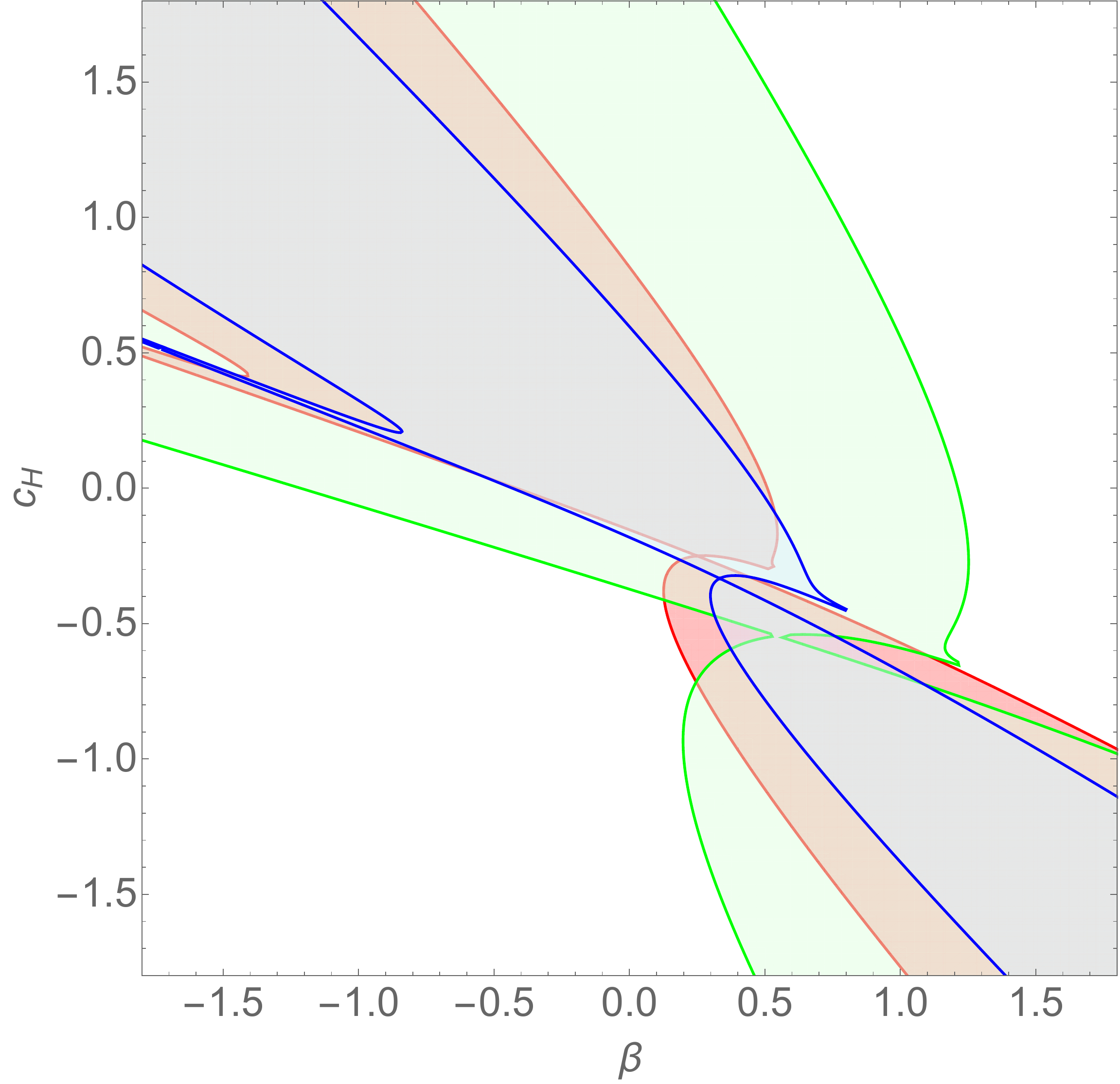}
     \caption{
     Same as Fig.~\ref{fig:plot_xis} but the condition $|\xi_{\rm t}-15/1144|\leq 0.2$.
	}
     \label{fig:plot_xit}
\end{figure}

Based on the resultant formulae, we now investigate expected constraints on the DHOST cosmology using current and future observations of large-scale structure. As discussed in Sec.~\ref{sec:Introduction}, the constraints on the growth index from the current observations have been already reported. The typical value of the observational error of the growth index in the current status is roughly estimated as $\lesssim 0.1$. Therefore, let us employ $\gamma =6/11\pm 0.1$ as a conservative constraint. On the other hand, to the best of our knowledge, no one puts the constraints on $E_{\rm f}$, $E_{\rm s}$ and $E_{\rm t}$ (or $\xi_{\rm f}$, $\xi_{\rm s}$ and $\xi_{\rm t}$) from observational data.
In this paper, as an empirical test, we assume that the error of the second-order indices is $\approx 0.2$. Namely, we consider $\xi_{\rm s}=0.0\pm 0.2$ and $\xi_{\rm t}=15/1144
\pm 0.2$. As for $\xi_{\rm f}$, it would be difficult to put a tighter constraint than that of $\gamma$ itself, since $\xi_{\rm f}=\gamma -\xi_\kappa$ [see Eq.~\eqref{eq:E_f exp}]. Hence, we focus only on $\xi_{\rm s}$ and $\xi_{\rm t}$ hereafter. Given the parameter set $(p,q)$, we can translate the constraints on $(c_{\rm H},\beta)$ using Eqs.~\eqref{eq:gamma pq} and \eqref{eq:xis pq}.

\begin{figure}
    \includegraphics[height=80mm]{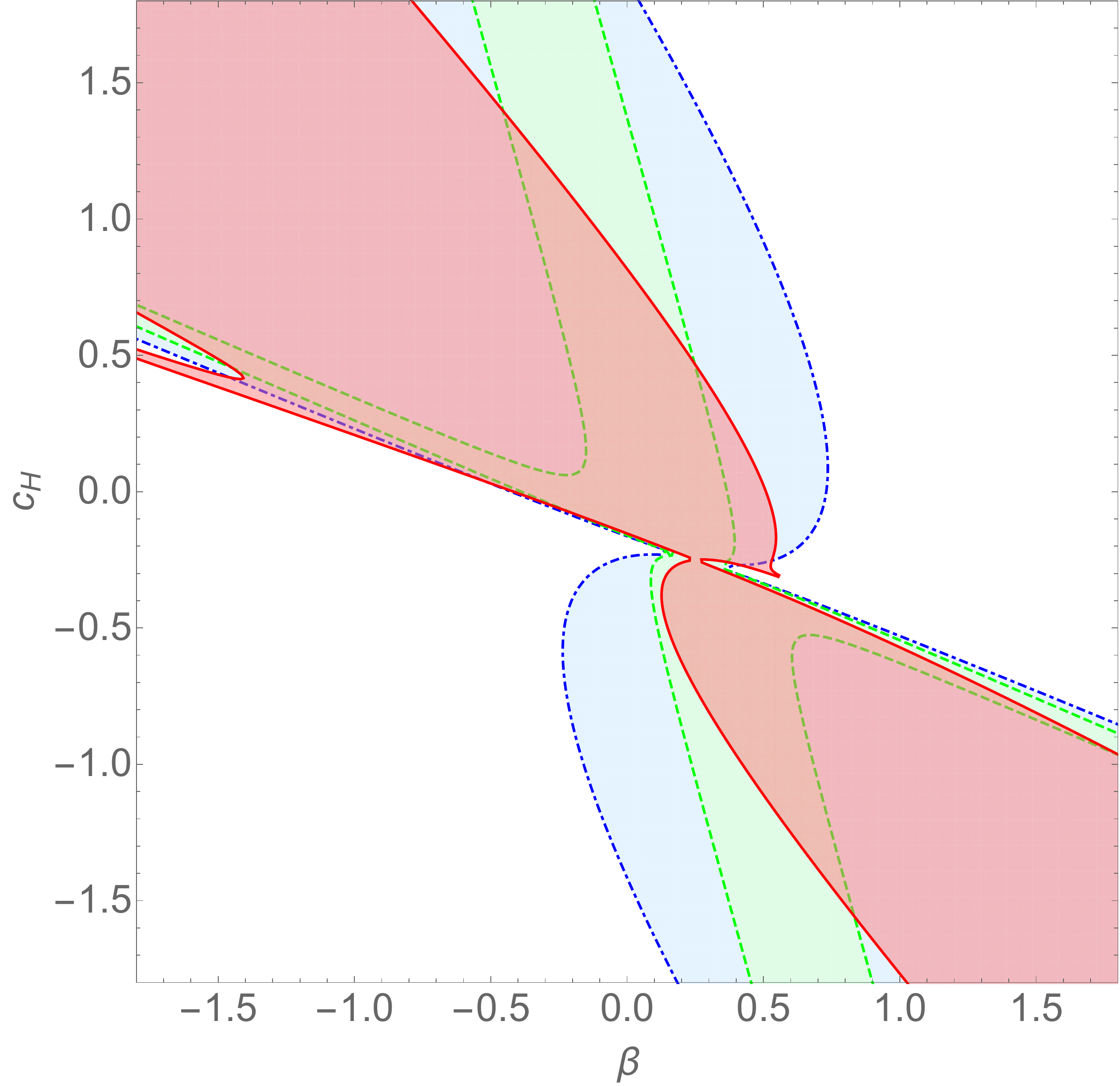}
     \caption{
     The allowed parameter region in $(\beta ,c_{\rm H})$ plane from the gravitational growth index 
     $|\gamma -6/11|\lesssim 0.1$ (green), the second-order index $|\xi_{\rm s}|\lesssim 0.2$ (blue), and
     $|\xi_{\rm t}-15/1144|\lesssim 0.2$ (red). The parameters used here are given by 
     $p=1$ and $q=1/2$.
	}
     \label{fig:plot_beta-cH}
\end{figure}

The parameter regions in the $(\beta,c_{\rm H})$ plane allowed by the constraints $|\xi_{\rm s}|\lesssim 0.2$ an $|\xi_{\rm t}-15/1144|\lesssim 0.2$ for the various values of $(p,q)$: $(1,1/2)$ (red), $(1,3/4)$ (blue), $(3,1)$ (green) are plotted in Figs.~\ref{fig:plot_xis} and \ref{fig:plot_xit}, respectively. These figures show that small changes in the parameters $(p,q)$ affect the details of the contour of the constant-$\xi_{\rm s}$ and constant-$\xi_{\rm t}$ curves, though generic features seem to remain unchanged. Hereafter, we focus on the specific parameter set $(p,q)=(1,1/2)$, corresponding to the model proposed in \cite{Crisostomi:2017pjs}. We show in Fig.~\ref{fig:plot_beta-cH} the allowed parameter region in the $(\beta,c_{\rm H})$ plane obtained from the constraints on the growth index $|\gamma -6/11|\lesssim 0.1$ (red), the second-order index $|\xi_{\rm s}|\lesssim 0.2$ (blue) and $|\xi_{\rm t}-15/1144|\lesssim 0.2$ (green). We found from Fig.~\ref{fig:plot_beta-cH} that the overlap region, where all these constraints are satisfied, is smaller than the individual allowed region. Therefore, the combined analysis is expected to provide tighter constraints on the DHOST cosmology only from the cosmological observations.

Before closing this section, we briefly discuss present-time observational bounds on the EFT parameters. The Newton potential $\Phi$ controls the stellar structure and is characterized by a combination of $\Upsilon_1=4[\alpha_{\rm H}+(1+\alpha_{\rm T})\beta_1]^2/[(1+\alpha_{\rm T})(1+\alpha_{\rm V}-4\beta_1)-\alpha_{\rm H}-1]$~\cite{Dima:2017pwp} (See also \cite{Langlois:2017dyl} for the case after GW170817). The lower bound has been obtained from the existence condition for stars in hydrostatic equilibrium $\Upsilon_1>-2/3$~\cite{Saito:2015fza}, while the upper bound comes from the comparison between the minimum mass of hydrogen-burning stars and the observed minimum red dwarf star $\Upsilon_1<1.6$~\cite{Sakstein:2015aac}. It is difficult to compare our results directly with these constraints since the leading order expression of $\gamma$, $\xi_{\rm s}$, and $\xi_{\rm t}$ are not necessarily valid all the way to the present time. Furthermore, assuming that $\alpha_{\rm H}=c_{\rm H}(1-\Omega_{\rm m})$, $\beta_1=\beta (1-\Omega_{\rm m})$, we can compare our results with the present-time bounds.

\section{Summary}
\label{sec:Summary}

In this paper, we have revisited the galaxy bispectrum as a possible probe to test the theory of gravity beyond linear-order perturbations and have discussed the potential impact of the second-order peculiar velocity field. We have derived the redshift-space galaxy bispectrum with the second-order kernels that include the effect of the modified gravity, and have shown that the signature of the modified gravity obtained from the kernel of the second-order density fluctuations is partially hidden by the uncertainty in the nonlinear galaxy bias functions. We also have pointed out that the contribution from the second-order peculiar velocity field in the galaxy bispectrum can be used to extract the higher-order properties of modified gravity without suffering from the uncertainty of the nonlinear galaxy bias function. Based on this fact, we have proposed the novel phenomenological parameters $E_{\rm f}=\Omega_{\rm m}^{\xi_{\rm f}}$, $E_{\rm s}=\Omega_{\rm m}^{\xi_{\rm s}}$ and $E_{\rm t}=\Omega_{\rm m}^{\xi_{\rm t}}$ [Eq.~\eqref{eq:E_s E_t def}] to trace the nonlinear growth history. We then have applied the formulae to the DHOST cosmology and found that the combined analysis of the growth rate and the second-order indices can provide a tight constraint on the DHOST cosmology.

We have developed the formulation of the time evolution for the first- and second-order density fluctuations in the framework of the DHOST theories. By expanding the background and perturbed equations for the density fluctuations in terms of $\varepsilon =1-\Omega_{\rm m}$, we have analytically obtained the expression of $\xi_{\rm f}$, $\xi_{\rm s}$ and $\xi_{\rm t}$ using the parameters that characterize the DHOST theories [Eqs.~\eqref{eq:xi_s} and \eqref{eq:E_t exp}], as well as the gravitational growth index $\gamma$ [Eq.~\eqref{eq:gamma exp}]. In particular, we have shown that the deviation of $E_{\rm s}$ from unity can be treated as a clear signal of the gravity theory describing the beyond-Horndeski theories.

Finally, as an application we have applied the resultant expressions to a specific cosmological model of the shift-symmetric DHOST theories, which has the observational constraint on the speed of gravitational waves. We then have obtained the new constraint on the parameter space and found that the analysis combined with the information obtained from $\gamma$, $\xi_{\rm s}$,
and $\xi_{\rm t}$ provides stringent constraints on the DHOST theories only from cosmological observations.

In this paper, we have considered only the leading term of $f$, $E_{\rm f}$, $E_{\rm s}$ and $E_{\rm t}$ as the asymptotic values in high redshifts. This assumption is expected to be valid for a wide class of modified gravity theories, and it would be interesting to investigate their time-evolution numerically. In addition, evaluating the galaxy bias function is important for the use of the galaxy bispectrum since the bias function may deviate from that of general relativity due to the modification of gravity theory. 
The information essentially required to constrain the parameters of $E_{\rm s}$ and $E_{\rm t}$ is the anisotropic component of the bispectrum in redshift space, which appears through the second-order velocity field. Therefore, the analysis of the anisotropic component will become more crucial in future bispectrum studies~\cite{Sugiyama:2018yzo,Sugiyama:2020uil}.

\acknowledgements

We thank Shun Arai, Tomohiro Fujita, Shin'ichi Hirano, and Tsutomu Kobayashi for 
the fruitful discussion.
This work was supported in part by JSPS KAKENHI Grant Nos.~17K14304 (D.Y.), 19H01891 (D.Y.), 19K14703 (N.S.S.).

\appendix

\section{Coefficients of first- and second-order solutions}
\label{sec:Coefficients of first- and second-order solutions}

In this Appendix, we briefly review the procedure used in Ref.~\cite{Hirano:2020dom}.

\subsection{First-order solutions}

To solve the perturbation equations, we first express the variables as
a perturbative series:
\al{
    \delta =\sum_{n=1}\delta_n
    \,,\ \ 
    \Phi =\sum_{n=1}\Phi_n
    \,,\cdots
}
where $\delta_n, \Phi_n,\cdots [={\cal O}(\delta_1^n)]$ denotes the $n$-th order quantities.
The equations-of-motion for the first-order $\Phi$ and $\Psi$ are schematically written as
\al{
    \mathsf{M}
    \left(\begin{array}{c}
    \Psi_1 \\ 
    \Phi_1 \\
    \end{array}\right) 
      = 
    \mathsf{N}
    \left(\begin{array}{c}
    Q_1 \\
    \dot Q_1/H \\
    \end{array}\right) 
    -\frac{a^2\rho_{\rm m}}{2M^2k^2}
    \left(\begin{array}{c}
    0 \\
    \delta_1 \\
     \end{array}\right) 
    \,,\label{eq:EOM for PsiPhi}
}
where $\mathsf{M}$ and $\mathsf{N}$ are $2\times 2$ matrix, which are explicitly defined as
\al{
    \mathsf{M} &
    =\left(\mathsf{M}\right)_{ab}
    =\left(\begin{array}{cc}
    \mathsf{M}_{\Psi\Psi} & \mathsf{M}_{\Psi\Phi} \\
    \mathsf{M}_{\Phi\Psi} & \mathsf{M}_{\Phi\Phi} \\
    \end{array}\right)
    = \left(\begin{array}{cc}
    1+\alpha_{\rm T} & -(1+\alpha_{\rm H}) \\
    1+\alpha_{\rm H} &-\beta_{3}/2 \\
    \end{array}\right)
    \,,\label{eq:M}
\\
    \mathsf{N} &
        =\left(\begin{array}{cc}
    \mathsf{N}_{\Psi Q} & \mathsf{N}_{\Psi \dot Q} \\
    \mathsf{N}_{\Phi Q} & \mathsf{N}_{\Phi \dot Q} \\
    \end{array}\right)
    = \left(\begin{array}{cc}
    \alpha_{\rm M}-\alpha_{\rm T}+\frac{\dot H}{H^2}\alpha_{\rm H} & -\alpha_{\rm H} \\
    \alpha_{\rm B}-\alpha_{\rm H}
	        -\frac{1}{aM^2}\left(\frac{aM^2\beta_1}{H}\right)^\cdot
	        +\frac{\dot H}{H^2}\beta_3 & -(2\beta_{1} +\beta_{3})/2 \\
    \end{array}\right)
    \,,\label{eq:N}
}
where the index $a,b$ stands for $\Psi$ and $\Phi$.
Moreover, when varying the effective Lagrangian with respect to $Q$, the equation-of-motion for the first-order scalar field perturbation is also 
schematically written as
\al{
	 \mathsf{R}_{\ddot Q}\frac{\ddot Q_1}{H^2}+\mathsf{R}_{\dot Q}\frac{\dot Q_1}{H} +\mathsf{R}_Q Q_1
		+\mathsf{R}_{\dot\Psi}\frac{\dot\Psi_1}{H} +\mathsf{R}_\Psi\Psi_1
		+\mathsf{R}_{\dot\Phi}\frac{\dot\Phi_1}{H} +\mathsf{R}_\Phi\Phi_1
		=0
	\,.\label{eq:EOM for Q}
}
The coefficients are given as
\al{
	&\mathsf{R}_\Psi =4\biggl[\alpha_{\rm M}-\alpha_{\rm T}+\frac{(aM^2\alpha_{\rm H})^\cdot}{aM^2H}\,\biggr]
	\,,\\
	&\mathsf{R}_\Phi =4\biggl[\alpha_{\rm B}-\alpha_{\rm H}+\frac{(aM^2\beta_3)^\cdot}{2aM^2H}\,\biggr]
	\,,\\
	&\mathsf{R}_{\dot\Psi}=4\alpha_{\rm H}
	\,,\\
	&\mathsf{R}_{\dot\Phi}=-2\left( 2\beta_1+\beta_3\right)
	\,,\\
	&\mathsf{R}_Q=2c_{QQ}+\frac{1}{aM^2}\left(\frac{aM^2(4\beta_1+\beta_3)}{H}\right)^{\cdot\cdot}
	\,,\\
	&\mathsf{R}_{\dot Q}=\frac{2H}{aM^2}\left(\frac{aM^2(4\beta_1+\beta_3)}{H^2}\right)^\cdot
	\,,\\
	&\mathsf{R}_{\ddot Q}=2\left( 4\beta_1+\beta_3\right)
	\,.
}
Let us solve Eqs.~\eqref{eq:EOM for PsiPhi} and \eqref{eq:EOM for Q} to express
$\Phi_1$, $\Psi_1$, and $Q_1$ in terms of $\delta_1$ and its time derivatives.
Solving Eq.~\eqref{eq:EOM for PsiPhi} for $\Phi_1$ and $\Phi_1$, and substituting into 
Eq.~\eqref{eq:EOM for Q}, one finds that the coefficients of $\ddot Q_1$ and $\dot Q_1$
become zero thanks to the degeneracy condition.
Hence, the first-order scalar perturbation $Q_1$ can be written in the form
\al{
    \frac{1}{a^2H^2}\pd^2Q_1=\kappa_Q\delta_1+\nu_Q\frac{\dot\delta_1}{H}
    \,,
}
where the coefficients can be written schematically as
\al{
    &\nu_{Q} =-\frac{3}{2}\frac{\Omega_{\rm m}}{Z}
        \Bigl[
		\mathsf{R}_{\dot\Psi}(\mathsf{M}^{-1})_{\Psi\Phi}+\mathsf{R}_{\dot\Phi}(\mathsf{M}^{-1})_{\Phi\Phi}
		\Bigr]
	\,,\\
	&\kappa_{Q} = -\frac{3}{2}\frac{\Omega_{\rm m}}{Z}
		\left[
		\mathsf{R}_\Psi (\mathsf{M}^{-1})_{\Psi\Phi} +\mathsf{R}_\Phi (\mathsf{M}^{-1})_{\Phi\Phi}  
		+\frac{aM^2}{H}\mathsf{R}_{\dot\Psi}\left[\frac{1}{aM^2}(\mathsf{M}^{-1})_{\Psi\Phi}\right]^{\cdot}  
		+\frac{aM^2}{H}\mathsf{R}_{\dot\Phi}\left[\frac{1}{aM^2}(\mathsf{M}^{-1})_{\Phi\Phi}\right]^{\cdot}
		\right]
	\,.
}
The denominator $Z$ is defined as
\al{
    Z &=\mathsf{R}_Q +\mathsf{R}_\Psi (\mathsf{M}^{-1}\mathsf{N})_{\Psi Q} 
			+\mathsf{R}_\Phi (\mathsf{M}^{-1}\mathsf{N})_{\Phi Q} 
			+\frac{1}{H}\mathsf{R}_{\dot\Psi}\left[(\mathsf{M}^{-1}\mathsf{N})_{\Psi Q}\right]^{\cdot} 
			+\frac{1}{H}\mathsf{R}_{\dot\Phi}\left[(\mathsf{M}^{-1}\mathsf{N})_{\Phi Q}\right]^{\cdot}
	\,.\label{eq:Z def}
}
Finally, substituting this back into Eq.~\eqref{eq:EOM for PsiPhi}, the first-order solutions
of the gravitational potentials $x_{1,a}=\{\Psi_1,\Phi_1\}$ can be expressed in terms of $\delta_1$, $\dot\delta_1$,
and $\ddot\delta_1$ as
\al{
    \frac{1}{a^2H^2}\pd^2 x_{1,a}=\kappa_a\delta_1+\nu_a\frac{\dot\delta_1}{H}+\mu_a\frac{\ddot\delta_1}{H^2}
    \,,
}
where the coefficients are written in terms of $\kappa_Q$ and $\nu_Q$
as well as the components of the matrices $\mathsf{M}$ and $\mathsf{N}$ by
\al{
    \mu_a &= (\mathsf{M}^{-1}\mathsf{N})_{a\dot Q}\,\nu_{Q}
    \,,\label{eq:mu_a def}\\
	\nu_a &= (\mathsf{M}^{-1}\mathsf{N})_{aQ}\,\nu_{Q}  
					+(\mathsf{M}^{-1}\mathsf{N})_{a\dot Q} \left[\kappa_{Q} +\frac{(a^{2}H\nu_{Q})^{\cdot}}{a^{2}H^{2}}\right]
	\,,\\
	\kappa_a &= \frac{3}{2}\Omega_{\rm m}(\mathsf{M}^{-1})_{a\Phi}  +(\mathsf{M}^{-1}\mathsf{N})_{aQ}\,\kappa_{Q}  
		+(\mathsf{M}^{-1}\mathsf{N})_{a\dot Q} \frac{(a^{2}H^{2}\kappa_{Q})^{\cdot}}{a^{2}H^{3}}
    \,.\label{eq:kappa_a def}
}

\subsection{Second-order solutions}

At the second-order, we need to take into account the contributions from 
the nonlinear mode-couplings.
To derive the second-order mode-couplings, it would be convenient
to introduce the first-order solutions as
\al{
    &K_a:=\frac{\pd^2x_{1,a}}{a^2H^2\delta_1}
    \,,\ \ 
    K_Q:=\frac{\pd^2Q_1}{a^2H^2\delta_1}
    \,,\ \ 
    K_{\dot Q}:=\frac{\pd^2\dot Q_1}{a^2H^3\delta_1}
    \,.\label{eq:K def}
}
Using the coefficients defined in the previous subsection, these can be rewritten as
\al{
    &K_\Phi 
        =\frac{\kappa_\Phi +f(\nu_\Phi -2\mu_\Phi )}{1-\mu_\Phi}
        =\frac{3}{2}\Omega_{\rm m}\Xi -f\varsigma
	\,,\\
	&K_\Psi =\kappa_\Psi +\frac{\mu_\Psi\kappa_\Phi}{1-\mu_\Phi}+f\left(\nu_\Psi -\frac{\mu_\Psi (2-\nu_\Phi)}{1-\mu_\Phi}\right)
	\,,\\
	&K_Q =\kappa_Q +f\nu_Q
	\,,\label{eq:K_Q def}\\
	&K_{\dot Q}=\frac{(a^2H^2D_+K_Q)^\cdot}{a^2H^3D_+}
	\,.
}
Then, the equations-of-motion for the second-order variables 
$\Psi_2$, $\Phi_2$, and $Q_2$ are schematically written as
\al{
    &\mathsf{M}
    \left(\begin{array}{c}
    \Psi_2 \\ 
    \Phi_2 \\
    \end{array}\right) 
      = 
    \mathsf{N}
    \left(\begin{array}{c}
    Q_2 \\
    \dot Q_2/H \\
    \end{array}\right) 
    -\frac{a^2\rho_{\rm m}}{2M^2k^2}
    \left(\begin{array}{c}
    0 \\
    \delta_2 \\
     \end{array}\right) 
    -\frac{a^2H^2}{k^2}\mathsf{O}
    \left(\begin{array}{c}
    W_{\alpha_{\rm s}} \\ 
    W_{\gamma} 
    \end{array}\right)
    \,,\label{eq:EOM for 2nd PsiPhi}\\
    &\mathsf{R}_{\ddot Q}\frac{\ddot Q_2}{H^2}+\mathsf{R}_{\dot Q}\frac{\dot Q_2}{H} +\mathsf{R}_Q Q_2
		+\mathsf{R}_{\dot\Psi}\frac{\dot\Psi_2}{H} +\mathsf{R}_\Psi\Psi_2
		+\mathsf{R}_{\dot\Phi}\frac{\dot\Phi_2}{H} +\mathsf{R}_\Phi\Phi_2
		=-\frac{a^2H^2}{k^2}\Bigl\{\mathsf{O}_{Q,\alpha}W_{\alpha_{\rm s}}+\mathsf{O}_{Q,\gamma}W_\gamma\Bigr\}
	\,.\label{eq:EOM for 2nd Q}
}
where $\mathsf{O}$ denotes the $2\times 2$ matrix characterizing
the amplitude of the nonlinear mode-couplings for the gravitational potentials, and
$\mathsf{O}_{Q,\Pi}$ represents the corresponding coefficient for $Q$ with
$\Pi$ representing the scale-dependence of the nonlinear mode-coupling.
We have defined $W_\Pi$ as
\al{
    W_\Pi ({\bm k})
        :=\int\frac{\dd^3{\bm p}_1\dd^3{\bm p}_2}{(2\pi)^3}
            \delta_{\rm D}^3({\bm k}-{\bm p}_1-{\bm p}_2)
            \Pi ({\bm p}_1,{\bm p}_2)
            \delta_{\rm L}({\bm p}_1)
            \delta_{\rm L}({\bm p}_2)
    \,.
}
Since $\Psi^{(2)}$ and $\Phi^{(2)}$ are generated by the scalar-scalar nonlinear interactions, 
as clearly seen in the effective Lagrangian Eq.~\eqref{eq:L3}, the mode-coupling coefficients 
$\mathsf{O}$ can be
expressed in terms of the first-order solution of the scalar field perturbation Eq.~\eqref{eq:K def} as
\al{
	\mathsf{O}
		=\left(
			\begin{array}{cc}
			\mathsf{O}_{\Psi,\alpha}& \mathsf{O}_{\Psi,\gamma} \\
			\mathsf{O}_{\Phi,\alpha} &  \mathsf{O}_{\Phi,\gamma} \\
			\end{array}
			\right)
		=\frac{1}{4}D_+^2K_Q^2
			\left(
			\begin{array}{cc}
			4\alpha_{\rm H} & \alpha_{\rm T} -4\alpha_{\rm H} \\
			2(2\beta_{1}+\beta_{3}) &  -\alpha_{\rm V}-\alpha_{\rm H}-2\beta_3 \\
			\end{array}
			\right)
	\,.\label{eq:O2}
}
On the other hand, the scalar-gravitational potential nonlinear interactions in addition to
the scalar-scalar nonlinear interactions can produce the nonlinear scalar perturbation $Q_2$.
Hence, $\mathsf{O}_{Q,\Pi}$ is written as
\al{
	\mathsf{O}_{Q,\alpha} 
		&= D_+^2K_Q\left\{
			4\alpha_{\rm H}K_{\Psi} 
			-2(2\beta_{1}+\beta_{3})K_{\Phi}
			+\mathsf{R}_{\dot Q}K_{Q}
			+6(4\beta_1+\beta_3)K_{\dot Q}
			\right\}
	\,,\\
	\mathsf{O}_{Q,\gamma} 
		&= -D_+^2K_Q\left\{
			2\alpha_{\rm T}K_{\Psi}
			+2(\alpha_{\rm V}-\alpha_{\rm H}-4\beta_1)K_{\Phi}
			+\left( c_{QQQ}+\mathsf{R}_{\dot Q}\right) K_{Q}
			+4(4\beta_{1}+\beta_{3})K_{\dot Q}
			\right\}
	\,.\label{eq:O_Q gamma}
}
Substituting the second-order solution of Eq.~\eqref{eq:EOM for 2nd PsiPhi} into Eq.~\eqref{eq:EOM for 2nd Q},
the second-order scalar perturbation $Q_2$ is expressed in the form
\al{
    \frac{1}{a^2H^2}\pd^2Q_2
            =\kappa_Q\delta_2+\nu_Q\frac{\dot\delta_2}{H}
                +D_+^2\left(\tau_{Q,\alpha}W_{\alpha_{\rm s}}+\tau_{Q,\gamma}W_\gamma\right)
    \,,
}
where the coefficients of the nonlinear mode-couplings with the shape $\Pi =\alpha_{\rm s},\gamma$ 
are given by
\al{
	&D_+^2\tau_{Q,\Pi} 
		=\frac{1}{Z}\biggl[
			\mathsf{O}_{Q,\Pi} 
			-\mathsf{R}_\Psi (\mathsf{M}^{-1}\mathsf{O})_{\Psi\Pi}
			-\mathsf{R}_\Phi (\mathsf{M}^{-1}\mathsf{O})_{\Phi\Pi}
	\notag\\
	&\quad\quad\quad\quad\quad\quad 
			-\frac{1}{a^{2}H^{3}}\mathsf{R}_{\dot\Psi}\left[ a^{2}H^{2}(\mathsf{M}^{-1}\mathsf{O})_{\Psi\Pi}\right]^{\cdot}
			-\frac{1}{a^{2}H^{3}}\mathsf{R}_{\dot\Phi}\left[ a^{2}H^{2}(\mathsf{M}^{-1}\mathsf{O})_{\Phi\Pi}\right]^{\cdot}
		\biggr]
	\,.\label{eq:tau_Q Pi}
}
Finally, substituting the solution back into Eq.~\eqref{eq:EOM for 2nd PsiPhi},
the second-order solutions of the gravitational potentials, $x_{2,a}=\{\Psi_2,\Phi_2\}$ are
\al{
    \frac{1}{a^2H^2}\pd^2 x_{2,a}=\kappa_a\delta_2 +\nu_a\frac{\dot\delta_2}{H}+\mu_a\frac{\ddot\delta_2}{H^2}
                +D_+^2\left(\tau_{a,\alpha}W_{\alpha_{\rm s}}+\tau_{a,\gamma}W_\gamma\right)
    \,,
}
with
\al{
	\tau_{a,\Pi} 
		&=(\mathsf{M}^{-1}\mathsf{O})_{a,\Pi}+(\mathsf{M}^{-1}\mathsf{N})_{aQ}\,\tau_{Q,\Pi} 
			+(\mathsf{M}^{-1}\mathsf{N})_{a\dot Q}\frac{(a^{2}H^{2}D_+^2\tau_{Q,\Pi})^{\cdot}}{a^{2}H^{3}D_+^2}
	\,.\label{eq:tau_a Pi}
}

\section{Explicit expression for some coefficients}
\label{sec:Explicit expression for some coefficients}

In this Appendix, we summarize some coefficients under the assumptions described in Sec.~\ref{sec:Setup}.
We first expand the time-dependent function $Z$ defined in Eq.~\eqref{eq:Z def} in terms of $\varepsilon =1-\Omega_{\rm m}$ as
\al{
    Z=Z^{(1)}\varepsilon +{\cal O}(\varepsilon^2)
    \,.
}
The leading coefficient is given by
\al{
	&Z^{(1)}=2\bigg\{
			3\left( 1+w^{(0)}\right) +2\left( c_{\rm M}-c_{\rm T}\right) 
			+\Bigl[ 1-2\left(c_{\rm M}-3w^{(0)}\right)\Bigr]\Bigl[ c_{\rm B}-c_{\rm H}-\beta\left( 1+c_{\rm M}-3w^{(0)}\right)\Bigr]
		\biggr\}
	\,.
}
Then, the expansion coefficients for $\Phi^{(1)}$, which are defined in Eqs.~\eqref{eq:mu_a def}--\eqref{eq:kappa_a def}, are 
\al{
    &\mu_\Phi^{(1)}
        =-\frac{6(c_{\rm H}+\beta)^2}{Z^{(1)}}
	\,,\label{eq:mu1}\\
	&\nu_\Phi^{(1)}
	        =-\frac{6(c_{\rm H}+\beta)^2(2+c_{\rm M}-3w^{(0)})}{Z^{(1)}}
	\,,\\
	&\kappa_\Phi^{(1)}
	       =-\frac{3}{2}\left(1+c_{\rm T}\right)+3\left( c_{\rm H}+\beta\right)
	           -\frac{6\rho^2}{Z^{(1)}}
				+\frac{3(c_{\rm H}+\beta)}{Z^{(1)}}
					\biggl\{
						\left( 3+2c_{\rm M}-6w^{(0)}\right)\rho
						-\left(c_{\rm M}-3w^{(0)}\right)\left( c_{\rm H}+\beta\right)
					\biggr\}
	\,.
}
with
\al{
    \rho =c_{\rm B}-c_{\rm M}+c_{\rm T}-\beta\left( c_{\rm M}-3w^{(0)}\right)
    \,.
}
We then obtain the explicit expression of the coefficients in Eqs.~\eqref{eq:mu exp}--\eqref{eq:Xi exp} as
\al{
	&\varsigma^{(1)}=\frac{6(c_{\rm H}+\beta )^2(c_{\rm M}-3w^{(0)})}{Z^{(1)}}
	\,,\label{eq:varsigma1}\\
	&\Xi^{(1)}=c_{\rm T}-2(c_{\rm H}+\beta)+\frac{4\rho^2}{Z^{(1)}}
				-\frac{2(c_{\rm H}+\beta)}{Z^{(1)}}
					\biggl\{
					    \left( 3+2c_{\rm M}-6w^{(0)}\right)\rho
						+\left( 2-c_{\rm M}+3w^{(0)}\right)\left( c_{\rm H}+\beta\right)
					\biggr\}
	\,.\label{eq:Xi1}
}
To describe the nonlinear mode-coupling terms, we need to expand the first-order solutions
Eq.~\eqref{eq:K def} as
\al{
    K_\Psi =\frac{3}{2}+{\cal O}(\varepsilon)
    \,,\ \ 
    K_\Phi =\frac{3}{2}+{\cal O}(\varepsilon)
    \,,\ \ 
    K_Q=K_Q^{(0)}+{\cal O}(\varepsilon)
    \,,\label{eq:K exp}
}
which implies that only the first-order scalar field perturbation has
the nontrivial zeroth-order contribution.
We note from the form of $K_Q$, Eq.~\eqref{eq:K_Q def} that $K_Q^{(0)}$ 
can be well described by
the combination of only the coefficients of the first-order equation-of-motion for $Q$, Eq.~\eqref{eq:EOM for Q}, as
\al{
    K_Q=\frac{\mathsf{R}_\Psi +\mathsf{R}_{\dot\Psi}+\mathsf{R}_\Phi +\mathsf{R}_{\dot\Phi}}{\mathsf{R}_Q}
        +{\cal O}(\varepsilon^2)
    \,.
}
The explicit form of the leading order coefficient is given as
\al{
    K_Q^{(0)} =&\frac{6\Bigl[\rho-\left(c_{\rm H}+\beta\right)\left( 2+c_{\rm M}-3w^{(0)}\right)\Bigr]}{Z^{(1)}}
    \,.
}
Substituting Eqs.~\eqref{eq:O2}--\eqref{eq:O_Q gamma} into Eq.~\eqref{eq:tau_Q Pi}
and expanding it in terms of $\varepsilon$, we have
\al{
	&\tau_{Q,\alpha}=\frac{6(c_{\rm H}+\beta)K_Q^{(0)}}{Z^{(1)}}+{\cal O}(\varepsilon)
	\,,\\
	&\tau_{Q,\gamma}=-\frac{K_Q^{(0)}}{Z^{(1)}}\biggl[ 
	        3\left( c_{\rm T}+c_{\rm V}-c_{\rm H}-4\beta\right)
	        +c_{QQQ}^{(1)}K_Q^{(0)}
	            \biggr]+{\cal O}(\varepsilon)
	\,,
}
with
\al{
    c_{QQQ}^{(1)}
            =4c_{\rm B}-2c_{\rm M}+3c_{\rm T}-3c_{\rm H}-c_{\rm V}+\left( c_{\rm V}-c_{\rm H}-8\beta\right)\left( c_{\rm M}-3w^{(0)}\right)
    \,.
}
Then, substituting them into Eq.~\eqref{eq:tau_a Pi}, we finally obtain
\al{
	&\tau_{\Phi ,\alpha}^{(1)}
		=\frac{3(c_{\rm H}+\beta)^2(7+2c_{\rm M}-6w^{(0)})K_Q^{(0)}}{Z^{(1)}}
	\,,\label{eq:tau_Phi,alpha}
}
and
\al{
	\tau_{\Phi ,\gamma}^{(1)}
		=&-\frac{1}{4}\left(K_Q^{(0)}\right)^2
		    \left(c_{\rm T}+c_{\rm V}-5c_{\rm H}-12\beta\right)
		    +\Bigl[\rho+\frac{3}{2}(c_{\rm H}+\beta)\Bigr]\tau_{Q,\gamma}
	\,.
}

\section{Different parametrization of fractional nonrelativistic matter density}
\label{sec:Diffrent parametrzation}

In this paper, we have treated the fractional nonrelativistic matter density $\Omega_{\rm m}$ as a time variable to evaluate the growth rate and the second-order variables analytically. When comparing our results with actual observational data, we need to evaluate $\Omega_{\rm m}$ as a function of redshift. However, in the framework of our formulation, it is not easy to solve $\Omega_{\rm m}$ explicitly. In this section, we discuss a possible prescription for replacing $\Omega_{\rm m}$ with the one defined in the standard $\Lambda$CDM Universe. The fractional energy density in the $\Lambda$CDM Universe is defined as
\al{
	&\Omega_{\rm m}^{\rm GR}=\frac{\rho_{\rm m}}{3M_{\rm Pl}^2H_{\rm GR}^2}
	\,,
}
where $M_{\rm Pl}$ denotes the Planck mass and $H_{\rm GR}^2$ is given by
the sum of the nonrelativistic matter and the cosmological constant $\Lambda$:
\al{
	H_{\rm GR}^2=\frac{1}{3M_{\rm Pl}^2}\rho_{\rm m}+\Lambda
	\,.
}
During the matter dominated era and the early stage of the dark energy dominated era, $\Omega_{\rm m}\approx 1$, we find that 
the time-evolution equation \eqref{eq:dotOmega_m} can reduce to
\al{
	\frac{\dd\Omega_{\rm m}}{\dd\ln a}\approx\left( 3w^{(0)}-c_{\rm M}\right)\left( 1-\Omega_{\rm m}\right)
	\,.
}
We can translate it into the equation for $\varepsilon\equiv 1-\Omega_{\rm m}$ as
\al{
	\frac{\dd\ln\varepsilon}{\dd\ln a}\approx 3w^{(0)}-c_{\rm M}
	\,.
}
We then solve it to obtain
\al{
	\varepsilon =\varepsilon (a_*)\left(\frac{a}{a_*}\right)^{-3w^{(0)}+c_{\rm M}}
	\,,
}
where $a_*$ denotes the initial time.
In the case of the $\Lambda$CDM Universe, the above expression becomes
\al{
	\varepsilon_{\rm GR}
	    \equiv 1-\Omega_{\rm m}^{\rm GR}
	    =\varepsilon_{\rm GR} (a_*)\left(\frac{a}{a_*}\right)^3
	\,.
}
Here, we take the deep matter dominated era as the initial time $a_*$, $a_*\ll 1$.
At the deep matter dominated era, we assume that the Universe can be well described 
by the $\Lambda$CDM Universe.
Namely, $\varepsilon (a_*)\approx\varepsilon_{\rm GR}(a_*)$.
Therefore, the difference between $\Omega_{\rm m}$ and $\Omega_{\rm m}^{\rm GR}$ can be well approximated as
\al{
	\Omega_{\rm m}-\Omega_{\rm m}^{\rm GR}
		\approx\Biggl[ 1-\left(\frac{a}{a_*}\right)^{-3(w^{(0)}+1)+c_{\rm M}}\,\Biggr]\left( 1-\Omega_{\rm m}^{\rm GR}\right)
	\,,
}
which immediately shows that the difference between $\Omega_{\rm m}$ and $\Omega_{\rm m}^{\rm GR}$ is suppressed by
$(1-\Omega_{\rm m}^{\rm GR})$.
During the matter dominated era and the early stage of the dark energy dominated era, $a_*\lesssim a\ll 1$,
the difference is further suppressed by the factor $(1-(a/a_*)^{-3(w^{(0)}+1)+c_{\rm M}})$, which is expected to be much smaller
than unity.
Hence, we expect that $\Omega_{\rm m}$ can be well approximated by $\Omega_{\rm m}^{\rm GR}$.



\begin{thebibliography}{99}

\bibitem{Linder:2007hg}
E.~V.~Linder and R.~N.~Cahn,
Astropart. Phys. \textbf{28} (2007), 481-488
doi:10.1016/j.astropartphys.2007.09.003
[arXiv:astro-ph/0701317 [astro-ph]].

\bibitem{Grieb:2016uuo}
J.~N.~Grieb \textit{et al.} [BOSS],
Mon. Not. Roy. Astron. Soc. \textbf{467} (2017) no.2, 2085-2112
doi:10.1093/mnras/stw3384
[arXiv:1607.03143 [astro-ph.CO]].

\bibitem{Sanchez:2016sas}
A.~G.~Sanchez \textit{et al.} [BOSS],
Mon. Not. Roy. Astron. Soc. \textbf{464} (2017) no.2, 1640-1658
doi:10.1093/mnras/stw2443
[arXiv:1607.03147 [astro-ph.CO]].

\bibitem{Gil-Marin:2018cgo}
H.~Gil-Mar\'\i{}n, J.~Guy, P.~Zarrouk, E.~Burtin, C.~H.~Chuang, W.~J.~Percival, A.~J.~Ross, R.~Ruggeri, R.~Tojerio and G.~B.~Zhao, \textit{et al.}
Mon. Not. Roy. Astron. Soc. \textbf{477} (2018) no.2, 1604-1638
doi:10.1093/mnras/sty453
[arXiv:1801.02689 [astro-ph.CO]].

\bibitem{Zhao:2018gvb}
G.~B.~Zhao, Y.~Wang, S.~Saito, H.~Gil-Mar\'\i{}n, W.~J.~Percival, D.~Wang, C.~H.~Chuang, R.~Ruggeri, E.~M.~Mueller and F.~Zhu, \textit{et al.}
Mon. Not. Roy. Astron. Soc. \textbf{482} (2019) no.3, 3497-3513
doi:10.1093/mnras/sty2845
[arXiv:1801.03043 [astro-ph.CO]].

\bibitem{Yamauchi:2017ibz}
D.~Yamauchi, S.~Yokoyama and H.~Tashiro,
Phys. Rev. D \textbf{96} (2017) no.12, 123516
doi:10.1103/PhysRevD.96.123516
[arXiv:1709.03243 [astro-ph.CO]].

\bibitem{Namikawa:2018erh}
T.~Namikawa, F.~R.~Bouchet and A.~Taruya,
Phys. Rev. D \textbf{98} (2018) no.4, 043530
doi:10.1103/PhysRevD.98.043530
[arXiv:1805.10567 [astro-ph.CO]].

\bibitem{Gil-Marin:2016wya}
H.~Gil-Mar\'\i{}n, W.~J.~Percival, L.~Verde, J.~R.~Brownstein, C.~H.~Chuang, F.~S.~Kitaura, S.~A.~Rodr\'\i{}guez-Torres and M.~D.~Olmstead,
Mon. Not. Roy. Astron. Soc. \textbf{465} (2017) no.2, 1757-1788
doi:10.1093/mnras/stw2679
[arXiv:1606.00439 [astro-ph.CO]].

\bibitem{Slepian:2016kfz}
Z.~Slepian, D.~J.~Eisenstein, J.~R.~Brownstein, C.~H.~Chuang, H.~Gil-Mar\'\i{}n, S.~Ho, F.~S.~Kitaura, W.~J.~Percival, A.~J.~Ross and G.~Rossi, \textit{et al.}
Mon. Not. Roy. Astron. Soc. \textbf{469} (2017) no.2, 1738-1751
doi:10.1093/mnras/stx488
[arXiv:1607.06097 [astro-ph.CO]].

\bibitem{Pearson:2017wtw}
D.~W.~Pearson and L.~Samushia,
Mon. Not. Roy. Astron. Soc. \textbf{478} (2018) no.4, 4500-4512
doi:10.1093/mnras/sty1266
[arXiv:1712.04970 [astro-ph.CO]].

\bibitem{Sugiyama:2018yzo}
N.~S.~Sugiyama, S.~Saito, F.~Beutler and H.~J.~Seo,
Mon. Not. Roy. Astron. Soc. \textbf{484} (2019) no.1, 364-384
doi:10.1093/mnras/sty3249
[arXiv:1803.02132 [astro-ph.CO]].


\bibitem{Sugiyama:2020uil}
N.~S.~Sugiyama, S.~Saito, F.~Beutler and H.~J.~Seo,
Mon. Not. Roy. Astron. Soc. \textbf{501} (2021) no.2, 2862-2896
doi:10.1093/mnras/staa3725
[arXiv:2010.06179 [astro-ph.CO]].

\bibitem{Langlois:2015cwa}
D.~Langlois and K.~Noui,
JCAP \textbf{02} (2016), 034
doi:10.1088/1475-7516/2016/02/034
[arXiv:1510.06930 [gr-qc]].

\bibitem{Crisostomi:2016czh}
M.~Crisostomi, K.~Koyama and G.~Tasinato,
JCAP \textbf{04} (2016), 044
doi:10.1088/1475-7516/2016/04/044
[arXiv:1602.03119 [hep-th]].

\bibitem{Achour:2016rkg}
J.~Ben Achour, D.~Langlois and K.~Noui,
Phys. Rev. D \textbf{93} (2016) no.12, 124005
doi:10.1103/PhysRevD.93.124005
[arXiv:1602.08398 [gr-qc]].

\bibitem{BenAchour:2016fzp}
J.~Ben Achour, M.~Crisostomi, K.~Koyama, D.~Langlois, K.~Noui and G.~Tasinato,
JHEP \textbf{12} (2016), 100
doi:10.1007/JHEP12(2016)100
[arXiv:1608.08135 [hep-th]].

\bibitem{Langlois:2018dxi}
D.~Langlois,
Int. J. Mod. Phys. D \textbf{28} (2019) no.05, 1942006
doi:10.1142/S0218271819420069
[arXiv:1811.06271 [gr-qc]].

\bibitem{Kobayashi:2019hrl}
T.~Kobayashi,
Rept. Prog. Phys. \textbf{82} (2019) no.8, 086901
doi:10.1088/1361-6633/ab2429
[arXiv:1901.07183 [gr-qc]].

\bibitem{Hirano:2020dom}
S.~Hirano, T.~Kobayashi, D.~Yamauchi and S.~Yokoyama,
Phys. Rev. D \textbf{102} (2020) no.10, 103505
doi:10.1103/PhysRevD.102.103505
[arXiv:2008.02798 [gr-qc]].

\bibitem{Desjacques:2016bnm}
V.~Desjacques, D.~Jeong and F.~Schmidt,
Phys. Rept. \textbf{733} (2018), 1-193
doi:10.1016/j.physrep.2017.12.002
[arXiv:1611.09787 [astro-ph.CO]].

\bibitem{Scoccimarro:1999ed}
R.~Scoccimarro, H.~M.~P.~Couchman and J.~A.~Frieman,
Astrophys. J. \textbf{517} (1999), 531-540
doi:10.1086/307220
[arXiv:astro-ph/9808305 [astro-ph]].

\bibitem{Takushima:2013foa}
Y.~Takushima, A.~Terukina and K.~Yamamoto,
Phys. Rev. D \textbf{89} (2014) no.10, 104007
doi:10.1103/PhysRevD.89.104007
[arXiv:1311.0281 [astro-ph.CO]].

\bibitem{Takushima:2015iha}
Y.~Takushima, A.~Terukina and K.~Yamamoto,
Phys. Rev. D \textbf{92} (2015) no.10, 104033
doi:10.1103/PhysRevD.92.104033
[arXiv:1502.03935 [gr-qc]].

\bibitem{Crisostomi:2019vhj}
M.~Crisostomi, M.~Lewandowski and F.~Vernizzi,
Phys. Rev. D \textbf{101} (2020) no.12, 123501
doi:10.1103/PhysRevD.101.123501
[arXiv:1909.07366 [astro-ph.CO]].

\bibitem{Lewandowski:2019txi}
M.~Lewandowski,
JCAP \textbf{08} (2020), 044
doi:10.1088/1475-7516/2020/08/044
[arXiv:1912.12292 [astro-ph.CO]].

\bibitem{Bouchet:1994xp}
F.~R.~Bouchet, S.~Colombi, E.~Hivon and R.~Juszkiewicz,
Astron. Astrophys. \textbf{296} (1995), 575
[arXiv:astro-ph/9406013 [astro-ph]].

\bibitem{Bernardeau:2001qr}
F.~Bernardeau, S.~Colombi, E.~Gaztanaga and R.~Scoccimarro,
Phys. Rept. \textbf{367} (2002), 1-248
doi:10.1016/S0370-1573(02)00135-7
[arXiv:astro-ph/0112551 [astro-ph]].

\bibitem{Horndeski:1974wa}
G.~W.~Horndeski,
Int. J. Theor. Phys. \textbf{10} (1974), 363-384
doi:10.1007/BF01807638

\bibitem{Deffayet:2011gz}
C.~Deffayet, X.~Gao, D.~A.~Steer and G.~Zahariade,
Phys. Rev. D \textbf{84} (2011), 064039
doi:10.1103/PhysRevD.84.064039
[arXiv:1103.3260 [hep-th]].

\bibitem{Kobayashi:2011nu}
T.~Kobayashi, M.~Yamaguchi and J.~Yokoyama,
Prog. Theor. Phys. \textbf{126} (2011), 511-529
doi:10.1143/PTP.126.511
[arXiv:1105.5723 [hep-th]].

\bibitem{Hirano:2018uar}
S.~Hirano, T.~Kobayashi, H.~Tashiro and S.~Yokoyama,
Phys. Rev. D \textbf{97} (2018) no.10, 103517
doi:10.1103/PhysRevD.97.103517
[arXiv:1801.07885 [astro-ph.CO]].

\bibitem{Langlois:2017mxy}
D.~Langlois, M.~Mancarella, K.~Noui and F.~Vernizzi,
JCAP \textbf{05} (2017), 033
doi:10.1088/1475-7516/2017/05/033
[arXiv:1703.03797 [hep-th]].

\bibitem{Dima:2017pwp}
A.~Dima and F.~Vernizzi,
Phys. Rev. D \textbf{97} (2018) no.10, 101302
doi:10.1103/PhysRevD.97.101302
[arXiv:1712.04731 [gr-qc]].

\bibitem{Bellini:2015wfa}
E.~Bellini, R.~Jimenez and L.~Verde,
JCAP \textbf{05} (2015), 057
doi:10.1088/1475-7516/2015/05/057
[arXiv:1504.04341 [astro-ph.CO]].

\bibitem{Kobayashi:2014ida}
T.~Kobayashi, Y.~Watanabe and D.~Yamauchi,
Phys. Rev. D \textbf{91} (2015) no.6, 064013
doi:10.1103/PhysRevD.91.064013
[arXiv:1411.4130 [gr-qc]].

\bibitem{Hirano:2019scf}
S.~Hirano, T.~Kobayashi and D.~Yamauchi,
Phys. Rev. D \textbf{99} (2019) no.10, 104073
doi:10.1103/PhysRevD.99.104073
[arXiv:1903.08399 [gr-qc]].

\bibitem{Hiramatsu:2020fcd}
T.~Hiramatsu and D.~Yamauchi,
Phys. Rev. D \textbf{102} (2020) no.8, 083525
doi:10.1103/PhysRevD.102.083525
[arXiv:2004.09520 [astro-ph.CO]].

\bibitem{Hirano:2019nkz}
S.~Hirano, T.~Kobayashi, D.~Yamauchi and S.~Yokoyama,
Phys. Rev. D \textbf{99} (2019) no.10, 104051
doi:10.1103/PhysRevD.99.104051
[arXiv:1902.02946 [astro-ph.CO]].

\bibitem{Crisostomi:2017pjs}
M.~Crisostomi and K.~Koyama,
Phys. Rev. D \textbf{97} (2018) no.8, 084004
doi:10.1103/PhysRevD.97.084004
[arXiv:1712.06556 [astro-ph.CO]].

\bibitem{TheLIGOScientific:2017qsa}
B.~P.~Abbott \textit{et al.} [LIGO Scientific and Virgo],
Phys. Rev. Lett. \textbf{119} (2017) no.16, 161101
doi:10.1103/PhysRevLett.119.161101
[arXiv:1710.05832 [gr-qc]].

\bibitem{Monitor:2017mdv}
B.~P.~Abbott \textit{et al.} [LIGO Scientific, Virgo, Fermi-GBM and INTEGRAL],
Astrophys. J. Lett. \textbf{848} (2017) no.2, L13
doi:10.3847/2041-8213/aa920c
[arXiv:1710.05834 [astro-ph.HE]].

\bibitem{Langlois:2017dyl}
D.~Langlois, R.~Saito, D.~Yamauchi and K.~Noui,
Phys. Rev. D \textbf{97} (2018) no.6, 061501
doi:10.1103/PhysRevD.97.061501
[arXiv:1711.07403 [gr-qc]].

\bibitem{Creminelli:2017sry}
P.~Creminelli and F.~Vernizzi,
Phys. Rev. Lett. \textbf{119} (2017) no.25, 251302
doi:10.1103/PhysRevLett.119.251302
[arXiv:1710.05877 [astro-ph.CO]].

\bibitem{Creminelli:2018xsv}
P.~Creminelli, M.~Lewandowski, G.~Tambalo and F.~Vernizzi,
JCAP \textbf{12} (2018), 025
doi:10.1088/1475-7516/2018/12/025
[arXiv:1809.03484 [astro-ph.CO]].

\bibitem{Creminelli:2019kjy}
P.~Creminelli, G.~Tambalo, F.~Vernizzi and V.~Yingcharoenrat,
JCAP \textbf{05} (2020), 002
doi:10.1088/1475-7516/2020/05/002
[arXiv:1910.14035 [gr-qc]].

\bibitem{Crisostomi:2019yfo}
M.~Crisostomi, M.~Lewandowski and F.~Vernizzi,
Phys. Rev. D \textbf{100} (2019) no.2, 024025
doi:10.1103/PhysRevD.100.024025
[arXiv:1903.11591 [gr-qc]].

\bibitem{Bahamonde:2019ipm}
S.~Bahamonde, K.~F.~Dialektopoulos, V.~Gakis and J.~Levi Said,
Phys. Rev. D \textbf{101} (2020) no.8, 084060
doi:10.1103/PhysRevD.101.084060
[arXiv:1907.10057 [gr-qc]].

\bibitem{Bahamonde:2019shr}
S.~Bahamonde, K.~F.~Dialektopoulos and J.~Levi Said,
Phys. Rev. D \textbf{100} (2019) no.6, 064018
doi:10.1103/PhysRevD.100.064018
[arXiv:1904.10791 [gr-qc]].

\bibitem{Saito:2015fza}
R.~Saito, D.~Yamauchi, S.~Mizuno, J.~Gleyzes and D.~Langlois,
JCAP \textbf{06} (2015), 008
doi:10.1088/1475-7516/2015/06/008
[arXiv:1503.01448 [gr-qc]].

\bibitem{Sakstein:2015aac}
J.~Sakstein,
Phys. Rev. D \textbf{92} (2015), 124045
doi:10.1103/PhysRevD.92.124045
[arXiv:1511.01685 [astro-ph.CO]].

\end{thebibliography}
\end{document}